\documentclass[letterpaper,english,reprint, aps]{revtex4-1}
\usepackage[T1]{fontenc}
\usepackage[latin9]{inputenc}
\usepackage{verbatim}
\usepackage{bm}
\usepackage{tipa}
\usepackage{tipx}
\usepackage{amsmath}
\usepackage{amssymb}
\usepackage{graphicx}



\usepackage{babel}
\begin{document}
\global\long\def\ket#1{\left| #1 \right\rangle }%
 
\global\long\def\mcH{{\mathcal{H}}}%
 
\global\long\def\lrp#1{\left( #1 \right)}%
 
\global\long\def\lrb#1{\left[ #1 \right]}%
 
\global\long\def\lrc#1{\left\{  #1 \right\}  }%
\global\long\def\hatU{\hat{U}}%
 
\global\long\def\cohU{\hat{U}^{\dagger}}%

\global\long\def\bra#1{\left\langle #1\right|}%
\global\long\def\avg#1{\left\langle #1 \right\rangle }%
 
\global\long\def\coa{\hat{a}^{\dagger}}%
 
\global\long\def\aoa{\hat{a}}%
 
\global\long\def\cob{\hat{b}^{\dagger}}%
 
\global\long\def\aob{\hat{b}}%
 
\global\long\def\coc{\hat{c}^{\dagger}}%
 
\global\long\def\aoc{\hat{c}}%
 
\global\long\def\cod{\hat{d}^{\dagger}}%
 
\global\long\def\aod{\hat{d}}%
\global\long\def\aogam{\hat{\gamma}}%
\global\long\def\cogam{\hat{\gamma}^{\dagger}}%
 
\global\long\def\hatD{\hat{D}}%
 
\global\long\def\hatH{\hat{H}}%
\global\long\def\hatN{\hat{N}}%
\global\long\def\aoA{\hat{A}}%
 
\global\long\def\aoB{\hat{B}}%
\global\long\def\hatQ{\hat{Q}}%
 
\global\long\def\bfalp{{\bf \mathbf{\alpha}}}%
\global\long\def\tr{\mathrm{Tr}}%
\global\long\def\half{\mathrm{\frac{1}{2}}}%
 
\global\long\def\tdpsi{\tilde{\psi}}%
\global\long\def\hatO{\hat{O}}%
\global\long\def\aor{\hat{\rho}}%
\global\long\def\hIO{\hat{O}^{I}}%
\global\long\def\hIH{\hat{H}^{I}}%
\global\long\def\bpsi{\bar{\psi}}%
\global\long\def\bphi{\bar{\phi}}%
\global\long\def\bPsi{\bar{\Psi}}%
\global\long\def\bPhi{\bar{\Phi}}%
\global\long\def\bvpsi{\breve{\psi}}%
\global\long\def\bvphi{\breve{\phi}}%
\global\long\def\bvbphi{\breve{\bphi}}%
\global\long\def\bvbpsi{\breve{\bpsi}}%
\global\long\def\barn{\bar{n}}%
\global\long\def\barT{\bar{T}}%
\global\long\def\barc{\bar{c}}%
\global\long\def\bart{\bar{t}}%
\global\long\def\barccT{\bar{\mathbb{T}}}%

\global\long\def\bflm{{\bf \text{\ensuremath{\bm{\lambda}}}}}%
\global\long\def\bfxi{{\bf \text{\ensuremath{\bm{\xi}}}}}%
\global\long\def\bfeta{\boldsymbol{\eta}}%
 
\global\long\def\bmlm{{\bf \text{\ensuremath{\bm{\lambda}}}}}%
\global\long\def\bfxi{{\bf \text{\ensuremath{\bm{\xi}}}}}%
\global\long\def\bmxi{{\bf \text{\ensuremath{\bm{\xi}}}}}%

\global\long\def\bfd{{\bf d}}%
\global\long\def\bfk{\bm{k}}%
\global\long\def\bfq{\bm{q}}%
\global\long\def\bmx{\bm{x}}%
\global\long\def\bmy{\bm{y}}%

\global\long\def\bfA{\bm{A}}%
\global\long\def\bfE{\bm{E}}%
 
\global\long\def\bfI{{\bf I}}%
\global\long\def\bfM{\bm{M}}%
\global\long\def\bfN{\bm{N}}%
\global\long\def\bmX{\bm{X}}%
\global\long\def\bmM{\bm{M}}%
\global\long\def\bmN{\bm{N}}%
\global\long\def\bmE{\bm{E}}%
\global\long\def\rmCAR{{\rm CAR}}%
\global\long\def\rmLAR{{\rm LAR}}%
\global\long\def\rmNT{{\rm NT}}%
\global\long\def\bmA{\bm{A}}%
\global\long\def\rmCT{{\rm CT}}%

\global\long\def\mcA{{\mathcal{A}}}%
 
\global\long\def\mcB{{\mathcal{B}}}%
 
\global\long\def\mcC{{\mathcal{C}}}%
 
\global\long\def\mcD{{\mathcal{D}}}%
 
\global\long\def\mcE{{\mathcal{E}}}%
 
\global\long\def\mcF{{\mathcal{F}}}%
 
\global\long\def\mcG{{\mathcal{G}}}%
 
\global\long\def\mcH{{\mathcal{H}}}%
 
\global\long\def\mcI{{\mathcal{I}}}%
 
\global\long\def\mcJ{{\mathcal{J}}}%
 
\global\long\def\mcK{{\mathcal{K}}}%
 
\global\long\def\mcL{{\mathcal{L}}}%
 
\global\long\def\mcM{{\mathcal{M}}}%
 
\global\long\def\mcN{\mathcal{N}}%
 
\global\long\def\mcO{{\mathcal{O}}}%
 
\global\long\def\mcP{{\mathcal{P}}}%
 
\global\long\def\mcQ{{\mathcal{Q}}}%
 
\global\long\def\mcR{{\mathcal{R}}}%
 
\global\long\def\mcS{{\mathcal{S}}}%
 
\global\long\def\mcT{\mathcal{T}}%
 
\global\long\def\mcU{{\mathcal{U}}}%
 
\global\long\def\mcV{{\mathcal{V}}}%
 
\global\long\def\mcW{{\mathcal{W}}}%
 
\global\long\def\mcX{{\mathcal{X}}}%
 
\global\long\def\mcY{{\mathcal{Y}}}%
 
\global\long\def\mcZ{{\mathcal{Z}}}%

\global\long\def\bbA{{\mathbb{A}}}%
 
\global\long\def\bbB{{\mathbb{B}}}%
 
\global\long\def\bbC{{\mathbb{C}}}%
 
\global\long\def\bbD{{\mathbb{D}}}%
 
\global\long\def\bbE{{\mathbb{E}}}%
 
\global\long\def\bbF{{\mathbb{F}}}%
 
\global\long\def\bbG{{\mathbb{G}}}%
 
\global\long\def\bbH{{\mathbb{H}}}%
 
\global\long\def\bbI{{\mathbb{I}}}%
 
\global\long\def\bbJ{{\mathbb{J}}}%
 
\global\long\def\bbK{{\mathbb{K}}}%
 
\global\long\def\bbL{{\mathbb{L}}}%
 
\global\long\def\bbM{{\mathbb{M}}}%
 
\global\long\def\bbN{{\mathbb{N}}}%
 
\global\long\def\bbO{{\mathbb{O}}}%
 
\global\long\def\bbP{{\mathbb{P}}}%
 
\global\long\def\bbQ{{\mathbb{Q}}}%
 
\global\long\def\bbR{{\mathbb{R}}}%
 
\global\long\def\bbS{{\mathbb{S}}}%
 
\global\long\def\bbT{\mathbb{T}}%
 
\global\long\def\bbU{{\mathbb{U}}}%
 
\global\long\def\bbV{{\mathbb{V}}}%
 
\global\long\def\bbW{{\mathbb{W}}}%
 
\global\long\def\bbX{{\mathbb{X}}}%
 
\global\long\def\bbY{{\mathbb{Y}}}%
 
\global\long\def\bbZ{{\mathbb{Z}}}%
\global\long\def\tdmcT{\mathcal{\tilde{T}}}%

\global\long\def\mfa{{\mathfrak{a}}}%
 
\global\long\def\mfb{{\mathfrak{b}}}%
 
\global\long\def\mfc{{\mathfrak{c}}}%
 
\global\long\def\mfd{{\mathfrak{d}}}%
 
\global\long\def\mfe{{\mathfrak{e}}}%
 
\global\long\def\mff{{\mathfrak{f}}}%
 
\global\long\def\mfg{{\mathfrak{g}}}%
 
\global\long\def\mfh{{\mathfrak{h}}}%
 
\global\long\def\mfi{{\mathfrak{i}}}%
 
\global\long\def\mfj{{\mathfrak{j}}}%
 
\global\long\def\mfk{{\mathfrak{k}}}%
 
\global\long\def\mfl{{\mathfrak{l}}}%
 
\global\long\def\mfm{{\mathfrak{m}}}%
 
\global\long\def\mfn{{\mathfrak{n}}}%
 
\global\long\def\mfo{{\mathfrak{o}}}%
 
\global\long\def\mfp{{\mathfrak{p}}}%
 
\global\long\def\mfq{{\mathfrak{q}}}%
 
\global\long\def\mfr{{\mathfrak{r}}}%
 
\global\long\def\mfs{{\mathfrak{s}}}%
 
\global\long\def\mft{{\mathfrak{t}}}%
 
\global\long\def\mfu{{\mathfrak{u}}}%
 
\global\long\def\mfv{{\mathfrak{v}}}%
 
\global\long\def\mfw{{\mathfrak{w}}}%
 
\global\long\def\mfx{{\mathfrak{x}}}%
 
\global\long\def\mfy{{\mathfrak{y}}}%
 
\global\long\def\mfz{{\mathfrak{z}}}%

\global\long\def\mfA{{\mathfrak{A}}}%
 
\global\long\def\mfB{{\mathfrak{B}}}%
 
\global\long\def\mfC{{\mathfrak{C}}}%
 
\global\long\def\mfD{{\mathfrak{D}}}%
 
\global\long\def\mfE{{\mathfrak{E}}}%
 
\global\long\def\mfF{{\mathfrak{F}}}%
 
\global\long\def\mfG{{\mathfrak{G}}}%
 
\global\long\def\mfH{{\mathfrak{H}}}%
 
\global\long\def\mfI{{\mathfrak{I}}}%
 
\global\long\def\mfJ{{\mathfrak{J}}}%
 
\global\long\def\mfK{{\mathfrak{K}}}%
 
\global\long\def\mfL{{\mathfrak{L}}}%
 
\global\long\def\mfM{{\mathfrak{M}}}%
 
\global\long\def\mfN{{\mathfrak{N}}}%
 
\global\long\def\mfO{{\mathfrak{O}}}%
 
\global\long\def\mfP{{\mathfrak{P}}}%
 
\global\long\def\mfQ{{\mathfrak{Q}}}%
 
\global\long\def\mfR{{\mathfrak{R}}}%
 
\global\long\def\mfS{{\mathfrak{S}}}%
 
\global\long\def\mfT{{\mathfrak{T}}}%
 
\global\long\def\mfU{{\mathfrak{U}}}%
 
\global\long\def\mfV{{\mathfrak{V}}}%
 
\global\long\def\mfW{{\mathfrak{W}}}%
 
\global\long\def\mfX{{\mathfrak{X}}}%
 
\global\long\def\mfY{{\mathfrak{Y}}}%
 
\global\long\def\mfZ{{\mathfrak{Z}}}%

\global\long\def\mrA{{\mathrm{A}}}%
 
\global\long\def\mrB{{\mathrm{B}}}%
 
\global\long\def\mrC{{\mathrm{C}}}%
 
\global\long\def\mrD{{\mathrm{D}}}%
 
\global\long\def\mrE{{\mathrm{E}}}%
 
\global\long\def\mrF{{\mathrm{F}}}%
 
\global\long\def\mrG{{\mathrm{G}}}%
 
\global\long\def\mrH{{\mathrm{H}}}%
 
\global\long\def\mrI{{\mathrm{I}}}%
 
\global\long\def\mrJ{{\mathrm{J}}}%
 
\global\long\def\mrK{{\mathrm{K}}}%
 
\global\long\def\mrL{{\mathrm{L}}}%
 
\global\long\def\mrM{{\mathrm{M}}}%
 
\global\long\def\mrN{{\mathrm{N}}}%
 
\global\long\def\mrO{{\mathrm{O}}}%
 
\global\long\def\mrP{{\mathrm{P}}}%
 
\global\long\def\mrQ{{\mathrm{Q}}}%
 
\global\long\def\mrR{{\mathrm{R}}}%
 
\global\long\def\mrS{{\mathrm{S}}}%
 
\global\long\def\mrT{{\mathrm{T}}}%
 
\global\long\def\mrU{{\mathrm{U}}}%
 
\global\long\def\mrV{{\mathrm{V}}}%
 
\global\long\def\mrW{{\mathrm{W}}}%
 
\global\long\def\mrX{{\mathrm{X}}}%
 
\global\long\def\mrY{{\mathrm{Y}}}%
 
\global\long\def\mrZ{{\mathrm{Z}}}%
\global\long\def\barbbC{\bar{\mathbb{C}}}%
\global\long\def\tdbbT{\tilde{\bbT}}%
\global\long\def\intw{{\bf \text{\ensuremath{\int\frac{d\omega}{2\pi}}}}}%

\global\long\def\msG{\mathscr{G}}%

\preprint{APS/123-QED}
\title{Microreversibility, fluctuation relations, and response properties
in 1D Kitaev Chain}
\author{Fan Zhang}
\affiliation{School of Physics, Peking University.}
\author{Jiayin Gu}
\affiliation{School of Physics, Peking University.}
\author{H. T. Quan}
\email{Corresponding author: htquan@pku.edu.cn}

\affiliation{School of Physics, Peking University, Beijing, 100871, China }
\affiliation{Collaborative Innovation Center of Quantum Matter, Beijing 100871,
China}
\affiliation{Frontiers Science Center for Nano-optoelectronics, Peking University,
Beijing, 100871, China}
\date{\today}
\begin{abstract}
We analytically calculate the cumulant generating function of energy
and particle transport in an open 1D Kitaev chain by utilizing the
Keldysh technique. The joint distribution of particle and energy currents
obeys different fluctuation relations in different regions of the
parameter space as a result of $U$(1) symmetry breaking and energy
conservation. We discuss the thermoelectricity of the Kitaev chain
as a three terminal system and derive an analytical expression of
the maximum work power. The response theory up to the second order
is explicitly checked, and the result is consistent with the relations
derived from the fluctuation relation.
\end{abstract}
\maketitle

\section{\label{sec:intro}Introduction}

Microreversibility, which is a fundamental symmetry of the physical
laws, imposes remarkable constraints on the nonequilibrium dynamics
of a system. The most famous example is the celebrated Onsager-Casimir
reciprocal relation, which states that the matrix of linear kinetic
coefficient is symmetric$\;$\citep{Onsager1931PR,Onsager1931PRa,Casimir1945RMP}.
This relation greatly reduces the number of response coefficients
in a transport process, thus finding wide applications in transport
experiments. Another example is the fluctuation-dissipation relation$\;$(FDR),
which relates the dissipation or response in a nonequilibrium process
to the properties in equilibrium$\;$\citep{Callen1951PR}. Recently,
a new family of nonequilibrium relations, called \emph{fluctuation
relations$\;$}(FR) have been discovered$\;$\citep{Evans1993prl,evans1994pre,Gallavotti1995jsp,Gallavotti1995prl,Kurchan1998jpa,Maes1999jsp,Lebowitz1999jsp,Esposito2009rmp,Campisi2011rmp,Jarzynski2011an,Seifert2012rpp}.
The derivation of the fluctuation relations only relies on the microreversibility
of the system, and does not depend on the microscopic details. These
fluctuation relations generalize the above two relations from the
linear-response regime to regimes arbitrarily far away from equilibrium.
From these relations, one can not only easily reproduce the results
in linear response theory such as the Green-Kubo formula, but also
obtain relations of higher-order response coefficients$\;$\citep{Saito2008prb,Andrieux2008prl,Andrieux2009njp,Gaspard2013njp,Barbier2018jpa,Gu2019pre,Barbier2020JPAMT,Barbier2020PRE,Gu2020jsm,Hu2022arxiv}.

The most general form of a fluctuation relation about the entropy
production in an open system can be written as
\[
\frac{P(\Delta\Omega)}{P(-\Delta\Omega)}=e^{\Delta\Omega},
\]
where $P(\Delta\Omega)$ is the probability distribution of entropy
production $\Delta\Omega$. According to principles of thermodynamics,
one can relate $\Delta\Omega$ to various physical observables, such
as particle number $N$, exchanged heat $Q_{h}$, and applied work
$W$. In an open system without driving, there is no work done on
the system. The entropy production is associated with the exchange
of particles and energy; the corresponding FR is termed as exchange
FR$\;$\citep{Jarzynski2004prl}.

In the derivation of exchange FR, two ingredients are used. One is
the microreversibility of the equation of motion and the other is
the particle and energy conservation. Whereas the former is well recognized,
the latter is implicit and taken for granted since the conservation
law is a result of $U(1)$ symmetry and time-translation symmetry.
However, in condensed matter physics, the $U(1)$ symmetry can be
explicitly broken in some systems described by low-energy effective
Hamiltonian, such as the BCS Hamiltonian of the superconductor. The
breaking of $U$(1) symmetry implies that the particle number is not
conserved in the transport process, and it can lead to new forms of
exchange FR.

In a previous work \citep{Zhang2021pre}, we show that the exchange
FR of particle current in the 1D Kitaev chain in the steady state
takes various forms for different parameters. It is due to the presence
of a paring term such as $\Delta\coc_{j+1}\coc_{j}$ which explicitly
breaks $U(1)$ symmetry. The competition between the paring potential,
the hopping amplitude, and the chemical potential gives rise to different
microscopic transport processes, namely, the normal transport$\;$(NT),
the local Andreev reflection$\;$(LAR) and the crossed Andreev reflection$\;$(CAR).
Each particle current component satisfies a steady-state FR. In this
article, we go one step further and study the joint probability distribution
of energy and particle transport. We use the Keldysh technique to
analytically calculate the full counting statistics of the particle
and energy currents. We will show that the joint distribution of particle
and energy currents obeys different exchange FRs due to $U(1)$ symmetry
breaking and energy conservation. We will study its response, linear
and nonlinear, and calculate the response coefficients. These response
coefficients are used to demonstrate a family of response relations
derived from the FR. In the linear response regime, we also discuss
the thermoelectricity of the Kitaev chain as a three terminal system.

Our paper is structured as follows. We introduce the open 1D Kitaev
chain model and analytically calculate its full counting statistics
of energy and particle in Sec. II. We discuss the exchange FR in Sec.
III. In Sec. IV, we study the response properties of the Kitaev chain.
In Sec. V, we discuss our results and make a summary.

\section{Model and Full Counting Statistics}

\begin{figure}
\includegraphics[scale=0.55]{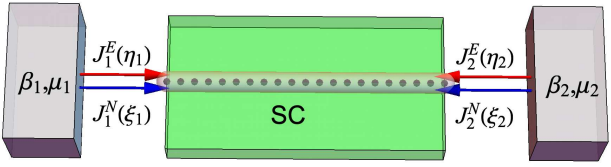}\caption{The setup of an open Kitaev chain. A nanowire is put above an s-wave
superconductor$\;$(SC) and couples to two reservoirs. The left and
right reservoirs are labelled as $\alpha=1,2$. The temperature and
chemical potential for reservoir 1$\;$(2) are $\beta_{1},\mu_{1}$$\;$($\beta_{2},\mu_{2}$),
respectively. The nonzero affinities will drive the system into a
steady state in the long-time limit. There are four currents: two
energy currents $J_{1}^{E},J_{2}^{E}$$\;$(red arrow) associated
with the counting fields $\eta_{1},\eta_{2}$, and two particle currents
$J_{1}^{N},J_{2}^{N}$$\;$(blue arrow) associated with the counting
fields $\xi_{1},\xi_{2}.$ \label{fig:The-setup}}
\end{figure}

We consider a Kitaev chain connected to two reservoirs. The set up
is shown in Fig.$\;$\ref{fig:The-setup}.
 A nanowire is put above an s-wave superconductor$\;$(SC) and couples
to two reservoirs. Due to the proximity effect, the Cooper pairs can
leak into the nanowire, and turns the low-energy effective Hamiltonian
of the nanowire into a 1D Kitaev chain$\;$\citep{Oreg2010prl,Lutchyn2010prl,qiao2021arxiv}.
The whole Hamiltonian is 
\begin{equation}
\hatH=\hatH_{{\rm K}}+\sum_{\alpha=1,2}\hatH_{\alpha}+\hatH_{{\rm I}},
\end{equation}
\begin{comment}
\[
\hatH=\hatH_{L}+\hatH_{K}+\hatH_{R}+\hatH_{LK}+\hatH_{KR}
\]
\end{comment}
where the Hamiltonian of the Kitaev chain is 
\begin{align}
\hatH_{{\rm K}} & =-\mu\sum_{j=1}^{N}\lrp{\coc_{j}\aoc_{j}-\half}\nonumber \\
 & \quad+\sum_{j=1}^{N-1}(-h~\coc_{j}\aoc_{j+1}+\Delta~\aoc_{j}\aoc_{j+1}+h.c)
\end{align}
with $\mu,h,\Delta$ the chemical potential, the hopping amplitude
and the superconducting gap. $\coc_{j}$ and $\aoc_{j}$ are the creation
and annihilation operators of electrons on site $j$; $N$ is the
site number of the Kitaev chain. The reservoirs are described by the
free fermion Hamiltonian 
\begin{equation}
\hatH_{\alpha}=\sum_{j}(\hbar\omega_{\alpha j}-\mu_{\alpha})\coc_{\alpha j}\aoc_{\alpha j},\quad\alpha=1,2.\label{eq:Ham_of_reservoirs}
\end{equation}
Here, $\omega_{\alpha j}$ denotes the energy of the $j$th state
of reservoir $\alpha$ whose chemical potential is $\mu_{\alpha}.$
We assume linear couplings between the Kitaev chain and the reservoirs
\begin{equation}
\hatH_{{\rm I}}=\sum_{j}\lambda_{1j}\lrp{\coc_{1j}\aoc_{1}+\coc_{1}\aoc_{1j}}+\sum_{j}\lambda_{2j}\lrp{\coc_{2j}\aoc_{N}+\coc_{N}\aoc_{2j}}
\end{equation}
with $\lambda_{\alpha j}$ the coupling strength. We adopt the two-point
measurement scheme. We assume that the initial state is a product
state, and every part is prepared in its thermal equilibrium state
$\hat{\rho}_{0}=e^{-\beta_{0}\hatH_{{\rm K}}-\sum_{\alpha=1,2}\beta_{\alpha}\hatH_{\alpha}}/\tr\lrb{e^{-\beta_{0}\hatH_{{\rm K}}-\sum_{\alpha=1,2}\beta_{\alpha}\hatH_{\alpha}}},$
where $\beta_{0}$ and $\beta_{\alpha}$ are the initial temperatures
of the Kitaev chain and reservoir $\alpha$, respectively. We measure
the particle number $\hatN_{\alpha}=\sum_{j}\coc_{\alpha j}\aoc_{\alpha j}$
and the energy $\hatH_{\alpha}$ of reservoir $\alpha$ simultaneously
at the initial time $t=0$ and a latter time $t=\tau.$ The particle$\;$(energy)
exchanged between the reservoir and the chain during a time interval
$[0,\tau]$ is defined to be the difference between the two outcomes
which is denoted as follows 
\begin{equation}
\Delta\bmX=\lrp{\Delta N_{1},\Delta N_{2},\Delta E_{1},\Delta E_{2}}=(\Delta\bmN,\Delta\bmE).
\end{equation}
We also define the operator $\hat{\bmX}=(\hatN_{1},\hatN_{2},\hatH_{1},\hatH_{2}).$
The moment generating function$\;$(MGF) which is the Fourier transform
of the probability distribution $P(\Delta\bmX)$ is defined as
\begin{align}
Z(\bmlm) & =\int d\Delta\bmX\;P(\Delta\bmX)e^{i\Delta\bfN\cdot\bfxi}e^{i\Delta\bfE\cdot\bfeta}\nonumber \\
 & =\tr\lrb{\hat{\rho}_{0}\cohU(\tau,0)e^{i\hat{\bmX}\cdot\bflm}\hat{U}(\tau,0)e^{-i\hat{\bmX}\cdot\bflm}}
\end{align}
with the counting fields $\bflm=(\bmxi,\bm{\eta})=(\xi_{1},\xi_{2},\eta_{1},\eta_{2})$
and $\hat{U}(\tau,0)$ the unitary evolution operator of the total
Hamiltonian.

We insert fermionic coherent states and write it in the form of contour
functional integral; by utilizing the Keldysh technique$\;$\citep{Kamenev2011book,Shankar2017QFT,Zhang2021pre},
we obtain the MGF of the open Kitaev chain in the long time limit
$\tau\to\infty.$
\begin{equation}
Z(\bmlm)=\prod_{\omega}\sqrt{Z(\bmlm;\omega)},
\end{equation}
 where the MGF of every mode is composed of three components: the
normal transport, the crossed Andreev reflection, and the local Andreev
reflection 
\begin{align}
Z(\bmlm;\omega) & =\frac{Z_{\rmNT}(\xi_{1}-\xi_{2})+Z_{\rmCAR}(\xi_{1}+\xi_{2})+Z_{\rmLAR}(\xi_{1},\xi_{2})}{Z_{\rmNT}(0)+Z_{\rmCAR}(0)+Z_{\rmLAR}(0)}.\label{eq:MGF}
\end{align}

\begin{widetext}
The three components are given by 
\begin{align*}
Z_{\rmNT}(\xi_{1}-\xi_{2}) & =\left\{ \bbC_{1}+\bbT_{1}\left[n_{1e}\barn_{2e}(e^{i(\xi_{1}-\xi_{2})}e^{i\omega\eta_{a}}-1)+\barn_{1e}n_{2e}(e^{-i(\xi_{1}-\xi_{2})}e^{-i\omega\eta_{a}}-1)\right]\right\} \\
 & \quad\times\left\{ \barbbC_{1}+\bar{\bbT}_{1}\left[n_{1h}\barn_{2h}(e^{-i(\xi_{1}-\xi_{2})}e^{i\omega\eta_{a}}-1)+\barn_{1h}n_{2h}(e^{i(\xi_{1}-\xi_{2})}e^{-i\omega\eta_{a}}-1)\right]\right\} ,\\
Z_{\rmCAR}(\xi_{1}+\xi_{2}) & =\left\{ \bbC_{2}+\bbT_{2}\left[n_{1e}\barn_{2h}(e^{i(\xi_{1}+\xi_{2})}e^{i\omega\eta_{a}}-1)+\barn_{1e}n_{2h}e^{-i(\xi_{1}+\xi_{2})}e^{-i\omega\eta_{a}}-1)\right]\right\} \\
 & \quad\times\left\{ \bar{\bbC}_{2}+\bar{\bbT}_{2}\left[n_{1h}\barn_{2e}(e^{-i(\xi_{1}+\xi_{2})}e^{i\omega\eta_{a}}-1)+\barn_{1h}n_{2e}(e^{i(\xi_{1}+\xi_{2})}e^{-i\omega\eta_{a}}-1)\right]\right\} ,\\
Z_{\rmLAR}(\xi_{1},\xi_{2}) & =\left\{ \bbC_{3}+\bbT_{3}\left[n_{1e}\barn_{1h}(e^{2i\xi_{1}}-1)+\barn_{1e}n_{1h}(e^{-2i\xi_{1}}-1)\right]\right\} \\
 & \quad\times\left\{ \bbC_{4}+\bbT_{4}\left[n_{2e}\barn_{2h}(e^{2i\xi_{2}}-1)+\barn_{2e}n_{2h}(e^{-2i\xi_{2}}-1)\right]\right\} .
\end{align*}
Here for simplicity, we eliminate the redundancy of the counting fields
of energy by introducing a new counting field $\eta_{a}=\eta_{1}-\eta_{2}$.
The fermionic occupation numbers of electrons and holes in reservoir
$\alpha$ are denoted by $n_{\alpha e}(\omega)=f_{\alpha}(\omega-\mu_{\alpha})$
and $n_{\alpha h}(\omega)=f_{\alpha}(\omega+\mu_{\alpha})$ respectively,
where $f_{\alpha}(\omega)=1/[e^{\beta_{\alpha}\omega}+1]$. $\barn_{\alpha e}(\omega)\equiv1-n_{\alpha e}(\omega)$
and $\barn_{\alpha h}(\omega)\equiv1-n_{\alpha h}(\omega).$ $\bbT_{j}(\omega)$
and $\bbC_{j}(\omega)$, $j=1,2,3,4$ are the transmission and reflection
amplitudes of mode $\omega$; $\bar{\bbT}_{j}(\omega)=\bbT_{j}(-\omega)$
and $\bar{\bbC}_{j}(\omega)=\bbC_{j}(-\omega).$We emphasize that
this form of MGF is valid for arbitrary number of sites and arbitrary
parameters. Different numbers of sites correspond to different amplitudes
$\bbC_{j}$ and $\bbT_{j}$ but do not affect the remaining expressions
of the MGF. In the following, we will only use $\eta_{a}$ as the
counting field for energy flow, i.e., $\bfeta=\eta_{a}$. The reduction
of the number of counting fields of energy is a consequence of energy
conservation. Note that the information of the initial state of the
Kitaev chain is lost in the long-time limit.
\end{widetext}

\section{Fluctuation Relation\label{sec:Fluctuation-Relation}}

From the explicit form of MGF, we observe that the transport process
is composed of independent bidirectional processes of mode $\omega$.
Every process consists of three subprocesses. The three subprocesses
are the NT, the CAR, and the LAR. The NT corresponds to transferring
one electron$\;$(hole) from the left reservoir to the right reservoir,
while the CAR corresponds to a process in which an incoming electron
from the left reservoir is turned into an outgoing hole in the right
reservoir$\;$\citep{Nilsson2008prl,Law2009prl}. As a result, one
electron from each reservoir is injected into the SC to form a Cooper
pair. The LAR corresponds to the process in which an incident electron
from one reservoir is converted into a backscattered hole. The CAR
and LAR break the particle conservation in two reservoirs, which is
due to the presence of a non-zero paring potential $\Delta$. If we
take $\Delta=0$, $\bbT_{2}$, $\bbT_{3}$ and $\bbT_{4}$ vanish
and only $\bbT_{1}$ is nonzero. In this case, the number of conservation
law recovers to two and only two counting fields $\xi_{a}\equiv\xi_{1}-\xi_{2}$,
$\eta_{a}$ are needed to generate the cumulants of the currents.
On the other hand, the particle conservation is also recovered if
we take into account the third reservoir, the superconductor, which
does not appear explicitly in the Hamiltonian of the open Kitaev chain.

\begin{figure}
\includegraphics[scale=0.6]{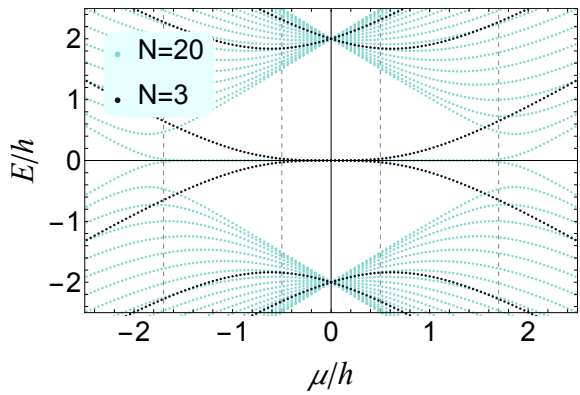}

\caption{The energy spectrum of a Kitaev chain with open boundary. Dark point
is for 3 sites; shallow point is for 20 sites. The regime for TSC
phase is reduced to about $|\mu/h|<0.5$ for a 3-site chain and about
$|\mu/h|<1.7$ for 20-site system; both are denoted as dashed vertical
lines. The phase boundary of a Kitaev chain in the thermodynamic limit
is $|\mu/h|<2.$ The shrink of the TSC phase is a manifestation of
the finite-size effect.\label{fig:The-energy-spectrum}}
\end{figure}

The MGF satisfies a symmetry relation
\begin{equation}
Z(\bfxi,\eta_{a})=Z(-\bfxi+i\bfA_{N},-\eta_{a}+iA_{E}),\label{eq:sym_in_CGF}
\end{equation}
 where $\bfA_{N}=(\beta_{1}\mu_{1},\beta_{2}\mu_{2})$ and $A_{E}=\beta_{2}-\beta_{1}$
are the affinities. The symmetry of the MGF implies an exchange FR
of the joint probability distribution$\;$\citep{Andrieux2009njp,Esposito2009rmp,Campisi2011rmp}
\begin{equation}
\frac{P(\Delta N_{1},\Delta N_{2},\Delta E_{1})}{P(-\Delta N_{1},-\Delta N_{2},-\Delta E_{1})}=e^{\Delta N_{1}\mu_{1}\beta_{1}+\Delta N_{2}\mu_{2}\beta_{2}}e^{\Delta E_{1}(\beta_{2}-\beta_{1})}.\label{eq:full_FR}
\end{equation}
Under certain conditions, one of the three current components dominates
the transport process and Eq.$\;$(\ref{eq:full_FR}) is reduced to
a simpler FR. We consider three different cases in the following.
The first case is when the pairing potential $\Delta=0$, i.e., the
Kitaev chain is a conventional conductor. The transmission amplitudes
of CAR $\bbT_{2}$ and LAR $\bbT_{3},\bbT_{4}$ vanish. The gain of
particles in one reservoir is equal to the loss of particles in the
other reservoir, i.e., $\Delta N_{1}=-\Delta N_{2}.$ The FR reads
\begin{equation}
\frac{P(\Delta N_{1},\Delta E_{1})}{P(-\Delta N_{1},-\Delta E_{1})}=e^{\Delta N_{1}(\mu_{1}\beta_{1}-\mu_{2}\beta_{2})}e^{\Delta E_{1}(\beta_{2}-\beta_{1})},\label{eq:FR_NT}
\end{equation}
which is the conventional FR of two terminal systems. If we introduce
a nonzero pairing potential $\Delta\neq0$ and turn off the hopping
term $h=0$, only CAR will occur$\;$\footnote{In fact, the CAR occurs when the site number is even. If the site
number is odd, NT rather than CAR will occur.}. In this case, $\Delta N_{1}=\Delta N_{2}$, and the FR becomes 
\begin{equation}
\frac{P(\Delta N_{1},\Delta E_{1})}{P(-\Delta N_{1},-\Delta E_{1})}=e^{\Delta N_{1}(\mu_{1}\beta_{1}+\mu_{2}\beta_{2})}e^{\Delta E_{1}(\beta_{2}-\beta_{1})}.\label{eq:FR_CAR}
\end{equation}
Two points are worth emphasizing. The first one is that when we apply
symmetric bias, i.e., $\mu_{1}\beta_{1}=-\mu_{2}\beta_{2},$ the probability
distribution $P(\Delta N_{1})$ is symmetric about $\Delta N_{1}=0$,
and gives zero mean particle current but nonzero energy current. The
second one is that when we apply equal bias, i.e., $\mu_{1}\beta_{1}=\mu_{2}\beta_{2},$
the nonzero particle current signatures the presence of a nonzero
paring potential. The third case is the Majorana case $\Delta=h$,
$\mu=0$. Two Majorana zero modes will emerge and localize at the
ends of the Kitaev chain. The NT and CAR are fully suppressed, that
is, $\bbT_{1}=\bbT_{2}=0$. The FR decouples 
\begin{equation}
\frac{P(\Delta N_{1})}{P(-\Delta N_{1})}=e^{\Delta N_{1}\mu_{1}\beta_{1}},\quad\frac{P(\Delta N_{2})}{P(-\Delta N_{2})}=e^{\Delta N_{2}\mu_{2}\beta_{2}}.\label{eq:FR_MZM}
\end{equation}
In the third case, there is no energy transport, since LAR effectively
transports two electrons with opposite energy to the chain. The net
exchange of particle is two, while the net exchange of energy is zero.
It is worth mentioning that the expression of transmission coefficient
of LAR $\bbT_{3(4)}$ in the third case is independent of the site
number$\;$\footnote{In fact, the number of sites should be larger than 3.}.
It can be proven that the Kitaev chain in the third case is equivalent
to a three-level system. It is also worth mentioning that the above
discussion of the MZM case applies when the localized MZM has no overlap,
i.e., the system is in the topological superconductor$\;$(TSC) phase.
For a infinite-long chain, the TSC phase appears when $|\mu/h|<2$
and $\Delta\neq0$$\;$\citep{kitaev2001ps}. For a short chain, the
TSC phase will shrink in the phase diagram due to the finite-size
effect$\;$(see Fig.$\;$\ref{fig:The-energy-spectrum}). 
In summary, all three subprocesses contribute to the particle current,
but only NT and CAR contribute to the energy flow.

\section{Response theory}

In this section, we first review the response theory which is derived
from the exchange FR$\;$\citep{Saito2008prb,Andrieux2009njp,Gaspard2013njp},
including the well-known results in linear response theory, such as
the Onsager reciprocal relation and FDR. Then we obtain the exact
expression of linear and nonlinear response coefficients of the Kitaev
chain and discuss thermoelectricity in this model.%
\begin{comment}
We will see that the particle-hole symmetry in the LAR make all the
high-order response coefficient vanish near the equilibrium, while
the CAR and NT have a finite contribution.
\end{comment}

\subsection{Response theory from FR}

The cumulants can be generated from the cumulant generating function$\;$(CGF)
which is defined as 
\begin{equation}
\mcF(\bflm,\bfA)\equiv\lim_{\tau\to\infty}\frac{1}{\tau}\ln Z(\bflm,\bfA),
\end{equation}
where we update the definition of counting fields $\bflm=\lrc{\lambda_{1},\lambda_{2},\lambda_{3}}\equiv\lrc{\xi_{1},\xi_{2},\eta_{a}}$.
The affinities are denoted as $\bfA=\{\bfA_{N},A_{E}\}.$ The CGF
inherits the symmetry in the MGF$\;${[}Eq.$\;$(\ref{eq:sym_in_CGF}){]},
namely, 
\begin{equation}
\mcF(\bflm,\bfA)=\mcF(-\bflm+i\bfA,\bfA).\label{eq:CGF_FR}
\end{equation}
All the cumulants of the energy and particle currents can be obtained
by successive derivatives with respect to the counting fields, and
then setting all these fields equal to zero. For example, the mean
value$\;$(first cumulant), the diffusivities$\;$(second cumulant),
and higher cumulants are given by
\begin{align}
J_{j}(\bfA) & \equiv\frac{\partial\mcF(\bmlm,\bfA)}{\partial(i\lambda_{j})}\bigg|_{\bmlm=0},\label{eq:current_FR}\\
D_{jk}(\bfA) & \equiv\half\frac{\partial^{2}\mcF(\bmlm,\bfA)}{\partial(i\lambda_{j})\partial(i\lambda_{k})}\bigg|_{\bmlm=0},\\
C_{jkl}(\bfA) & \equiv\frac{\partial^{3}\mcF(\bmlm,\bfA)}{\partial(i\lambda_{j})\partial(i\lambda_{k})\partial(i\lambda_{l})}\bigg|_{\bmlm=0}.\\
 & \ldots\nonumber 
\end{align}
As Refs.$\;$\citep{Saito2008prb,Andrieux2009njp,Gaspard2013njp}
point out, the third cumulants $C_{ijk}$ characterizes the magnetic-field
asymmetry of the fluctuations. When there is no term breaking time-reversal
symmetry, that is, no magnetic field, $C_{jkl}=0$.

At equilibrium$\;(\bfA=0)$, the mean current vanishes. We notice
that the mean current can be expanded in powers of the affinities
close to equilibrium
\[
J_{j}=\sum_{k}L_{j,k}A_{k}+\half\sum_{k,l}M_{j,kl}A_{k}A_{l}+\ldots.
\]
This expansion implies a definition of the response coefficients
\begin{align}
L_{j,k} & \equiv\frac{\partial^{2}\mcF(\bmlm,\bfA)}{\partial(i\lambda_{j})\partial A_{k}}\bigg|_{\substack{\bmlm=0\\
\bmA=0
}
},\label{eq:linear_response_def-2}\\
M_{j,kl} & \equiv\frac{\partial^{3}\mcF(\bmlm,\bfA)}{\partial(i\lambda_{j})\partial A_{k}\partial A_{l}}\bigg|_{\substack{\bmlm=0\\
\bmA=0
}
},\\
\vdots\nonumber 
\end{align}
The response coefficients and the cumulants satisfy a family of universal
relations, which can be derived from the exchange FR$\;$(\ref{eq:CGF_FR})$\;$\citep{Saito2008prb,Andrieux2009njp,Gaspard2013njp}.
The first-order response relations are nothing but the FDR 
\begin{equation}
L_{j,k}=D_{jk}(\bmA=0),\label{eq:FDR}
\end{equation}
and the Onsager reciprocal relation
\[
L_{k,j}=L_{j,k},
\]
where the second relation is from the symmetry of $D_{jk}=D_{kj}$.
Thus, we see that the two main cornerstones of linear response theory
are encoded in the exchange FR. 

As for nonlinear response at equilibrium, we have similar relations$\;$\citep{Saito2008prb,Andrieux2009njp,Gaspard2013njp}
\begin{align}
M_{i,jk} & =\lrp{\frac{\partial D_{ij}}{\partial A_{k}}+\frac{\partial D_{ik}}{\partial A_{j}}}\bigg|_{\bfA=0},\label{eq:sec_order_response_relation}\\
N_{i,jkl} & =\lrp{\frac{\partial^{2}D_{ij}}{\partial A_{k}\partial A_{l}}+\frac{\partial^{2}D_{ik}}{\partial A_{j}\partial A_{l}}+\frac{\partial^{2}D_{il}}{\partial A_{j}\partial A_{k}}-\frac{1}{2}\frac{\partial C_{ijk}}{\partial A_{l}}}\bigg|_{\bfA=0},\nonumber \\
 & \vdots\nonumber 
\end{align}
where $N_{i,jkl}$ is the third-order response coefficient
\[
N_{i,jkl}=\frac{\partial^{3}J_{i}(\bfA)}{\partial A_{j}\partial A_{k}\partial A_{l}}\bigg|_{\bmA=0}.
\]
In the following, we will consider response properties up to the second
order.

\subsection{Linear response in Kitaev chain}

From Eq.$\;$(\ref{eq:current_FR}), we obtain the expression of the
particle current $J_{1}^{N}$ from the left reservoir and $J_{2}^{N}$
from the right reservoir, as well as the energy current $J_{1}^{E}$
from the left reservoir
\begin{align}
J_{1}^{N} & =\int\frac{d\omega}{2\pi}\left[\tdbbT_{1}\lrp{p_{e\to e}-p_{e\gets e}}+\tdbbT_{2}\lrp{p_{e\to h}-p_{e\gets h}}\right.\nonumber \\
 & \quad\quad\quad\quad\left.+\tdbbT_{3}\lrp{n_{1e}\barn_{1h}-\barn_{1e}n_{1h}}\right],\label{eq:current_1N}\\
J_{2}^{N} & =\int\frac{d\omega}{2\pi}\left[\tdbbT_{1}\lrp{p_{e\gets e}-p_{e\to e}}+\tdbbT_{2}\lrp{p_{e\to h}-p_{e\gets h}}\right.\nonumber \\
 & \quad\quad\quad\quad\left.+\tdbbT_{4}\lrp{n_{2e}\barn_{2h}-\barn_{2e}n_{2h}}\right],\label{eq:current_2N}\\
J_{1}^{E} & =\int\frac{d\omega}{2\pi}\omega\lrb{\tdbbT_{1}\lrp{p_{e\to e}-p_{e\gets e}}+\tdbbT_{2}\lrp{p_{e\to h}-p_{e\gets h}}},\label{eq:current_1E}
\end{align}
\begin{comment}
\begin{align*}
J_{1E} & =\int\frac{d\omega}{2\pi}\omega\lrb{T_{41}\lrp{p_{e\to e}-p_{e\gets e}}+T_{42}\lrp{p_{e\to h}-p_{e\gets h}}},\\
J_{1N} & =\int\frac{d\omega}{2\pi}\left[T_{41}\lrp{p_{e\to e}-p_{e\gets e}}+T_{42}\lrp{p_{e\to h}-p_{e\gets h}}\right.\\
 & \quad\quad\quad\quad\left.T_{43}\lrp{n_{1e}\barn_{1h}-\barn_{1e}n_{1h}}\right],\\
J_{2N} & =\int\frac{d\omega}{2\pi}\left[T_{41}\lrp{p_{e\gets e}-p_{e\to e}}+T_{42}\lrp{p_{e\to h}-p_{e\gets h}}\right.\\
 & \quad\quad\quad\quad\left.T_{44}\lrp{n_{2e}\barn_{2h}-\barn_{2e}n_{2h}}\right],
\end{align*}
\end{comment}
where $p_{e\to e}=n_{1e}\barn_{2e}$, $p_{e\gets e}=\barn_{1e}n_{2e}$,
$p_{e\to h}=n_{1e}\barn_{2h},$ and $p_{e\gets h}=\barn_{1e}n_{2h}.$
The transmission coefficients are given by $\tdbbT_{1,2}=\barbbC_{1,2}\bbT_{1,2}/\bbC\bar{\bbC}$
with $\bbC\bar{\bbC}=\bbC_{1}\barbbC_{1}+\bbC_{2}\barbbC_{2}+\bbC_{3}\bbC_{4}.$
The energy current of the right reservoir $J_{2}^{E}$ is equal to
the opposite of $J_{1}^{E}$. Three independent currents are consistent
with three affinities. As mentioned before, the energy current is
carried by the particles participating in the NT and the CAR processes,
while all the three transport processes contribute to the particle
flow. From Eq.$\;$(\ref{eq:linear_response_def-2}), we obtain the
linear response coefficients $L_{\alpha,\beta}$ relevant to $J_{1}$
as$\;$(For simplicity, we label the currents ($J_{1}^{N},J_{2}^{N},J_{1}^{E}$)
as $(J_{1},J_{2},J_{3})$)
\begin{align}
L_{1,1} & =\int\frac{d\omega}{2\pi}\lrb{\tdbbT_{1}(\omega)+\tdbbT_{2}(\omega)+2\tdbbT_{3}(\omega)}n_{1e}\barn_{1e},\label{eq:linear_coeff_L11}\\
L_{1,2} & =\int\frac{d\omega}{2\pi}\lrb{-\tdbbT_{1}(\omega)n_{2e}\barn_{2e}+\tdbbT_{2}(\omega)n_{2h}\barn_{2h}},\label{eq:linear_coeff_L12}\\
L_{1,3} & =\int\frac{d\omega}{2\pi}\omega\left[\tdbbT_{1}(\omega)+\tdbbT_{2}(\omega)+2\tdbbT_{3}(\omega)\right]n_{1e}\barn_{1e}.\label{eq:linear_coeff_L13}
\end{align}
 The linear response coefficients $L_{\alpha,\beta}$ relevant to
$J_{2}$ are
\begin{align}
L_{2,1} & =\int\frac{d\omega}{2\pi}\lrb{-\tdbbT_{1}(\omega)+\tdbbT_{2}(\omega)}n_{1e}\barn_{1e},\label{eq:linear_coeff_L21}\\
L_{2,2} & =\int\frac{d\omega}{2\pi}\lrb{\tdbbT_{1}(\omega)+\tdbbT_{2}(-\omega)+2\tdbbT_{4}(\omega)}n_{2e}\barn_{2e},\label{eq:linear_coeff_L22}\\
L_{2,3} & =\int\frac{d\omega}{2\pi}\omega\left[-\tdbbT_{1}(\omega)+\tdbbT_{2}(\omega)\right]n_{1e}\barn_{1e}.\label{eq:linear_coeff_L23}
\end{align}
The linear response coefficients $L_{\alpha,\beta}$ relevant to $J_{3}$
are
\begin{align}
L_{3,1} & =\int\frac{d\omega}{2\pi}\omega\lrb{\tilde{\bbT}_{1}(\omega)+\tilde{\bbT}_{2}(\omega)}n_{1e}\barn_{1e},\label{eq:linear_coeff_L31}\\
L_{3,2} & =\int\frac{d\omega}{2\pi}\omega\lrb{-\tilde{\bbT}_{1}(\omega)n_{2e}\barn_{2e}+\tilde{\bbT}_{2}(\omega)n_{2h}\barn_{2h}},\label{eq:linear_coeff_L32}\\
L_{3,3} & =\int\frac{d\omega}{2\pi}\omega^{2}\lrb{\tilde{\bbT}_{1}(\omega)+\tilde{\bbT}_{2}(\omega)}n_{1e}\barn_{1e}.\label{eq:linear_coeff_L33}
\end{align}
From Eq.$\;$(\ref{eq:linear_coeff_L13}) and Eq.$\;$(\ref{eq:linear_coeff_L31}),
it seems that Onsager reciprocal relation is apparently violated due
to the presence of LAR. But actually, we have $\tdbbT_{3}(\omega)=\tdbbT_{3}(-\omega)$
as a consequence of the particle-hole symmetry, so the term containing
$\tdbbT_{3}$ in Eq.$\;$(\ref{eq:linear_coeff_L13}) is an odd function
of $\omega$ at zero affinity and vanishes after the integration.
Hence, the Onsager reciprocal relation remains valid in our model.

In experiment, the more familiar linear response coefficients are
the electrical conductance $G^{e}$, thermal conductance $K$, and
Seebeck coefficient $S$. In previous studies on the thermoelectricity
of 1D Kitaev chain or two Majorana zero modes$\;$(MZMs), the system
has been treated as a two-terminal system$\;$\citep{Lopez2014PRB,Ramos-Andrade2016PRB}.
Landauer-B$\ddot{{\rm u}}$ttiker formula is invoked to obtain the
currents, such as Eqs.$\;$(3,4) in Ref.$\;$\citep{Lopez2014PRB}$\;$(in
our notation)
\begin{align}
J_{1}^{N} & =\intw\;\bbT(\omega)[n_{1e}(\omega)-n_{2e}(\omega)],\label{eq:wrong_current}\\
J_{1}^{E} & =\intw\;\omega\bbT(\omega)[n_{1e}(\omega)-n_{2e}(\omega)].\label{eq:wrong_energy_current}
\end{align}
The above expression of particle current neglects the fact that there
are in total three current components. In Appendix$\;$\ref{sec:Two-Terminal-Majorana-Junction},
we show that Eq.$\;$(\ref{eq:wrong_current}) differs from Eq.$\;$(\ref{eq:current_1N_MZM})
obtained in the framework of FCS, and thus gives half the quantized
electrical conductance $e^{2}/h$$\;$(the correct one is $2e^{2}/h$).
The expression of energy current Eq.$\;$(\ref{eq:wrong_energy_current})
is also incorrect in the transmission term $\bbT(\omega)$, due to
the fact that $\bbT(\omega)$ incorrectly includes contributions from
LAR which is not involved in the energy transport. The correct expression
of currents and a detailed discussion of two MZMs coupled to two reservoirs
as a three-terminal system are given in Appendix$\;$\ref{sec:Two-Terminal-Majorana-Junction}.

In fact, the presence of three independent currents$\;$(and three
affinities) in this model implies that it is a genuine three-terminal
system, where the third terminal is the grounded SC. It can be compared
to the phonon-thermoelectric systems, e.g., a double quantum dots$\;$(QDs)
in contact with two metals and a phonon substrate$\;$\citep{Jiang2012PRB,Jiang2013NJP,Mazza2014NJP,Jiang2015PRB}.
The phonon bath absorbs or releases heat but does not exchange particles
with the two QDs, while the grounded SC in our system exchanges Cooper
pairs with the nanowire but does not exchange energy.

In a three-terminal system, we write the relation between the currents
and affinities in the linear response regime as 
\[
\begin{pmatrix}\begin{array}{c}
J_{1}^{N}\\
J_{2}^{N}\\
J_{1}^{Q}
\end{array}\end{pmatrix}=\begin{pmatrix}L_{1,1} & L_{1,2} & L_{1,3}\\
L_{2,1} & L_{2,2} & L_{2,3}\\
L_{3,1} & L_{3,2} & L_{3,3}
\end{pmatrix}\begin{pmatrix}\delta\mu_{1}/T\\
\delta\mu_{2}/T\\
\delta T/T^{2}
\end{pmatrix},
\]
where we take the temperature of the right reservoir as the reference
temperature $T=T_{2}$ and $\delta T=T_{1}-T_{2}$. The chemical potential
of the grounded SC $\mu_{{\rm SC}}=0$ is taken as the reference of
chemical potential and $\delta\mu_{\alpha}\equiv\mu_{\alpha}-\mu_{{\rm SC}}=\mu_{\alpha}$.
The heat current from the left reservoir is defined as $J_{1}^{Q}\equiv J_{1}^{E}-\mu_{1}J_{1}^{N}$
as a result of thermodynamical laws$\;$\citep{Whitney2018book}.
Following Ref.$\;$\citep{Mazza2014NJP}, the electrical conductance
is obtained under the isothermal condition, i.e.,
\begin{equation}
G_{ij}=\left(\frac{e^{2}J_{i}^{N}}{\delta\mu_{j}}\right)_{\substack{\delta T=0\\
\delta\mu_{k}=0\text{\;}k\neq i
}
}=\frac{e^{2}}{T}\begin{pmatrix}L_{1,1} & L_{1,2}\\
L_{2,1} & L_{2,2}
\end{pmatrix}.\label{eq:ele_cond_three}
\end{equation}
Here, $G_{11}$ and $G_{22}$ are the local electrical conductances
and $G_{12}(=G_{21})$ is the non-local electrical conductance. The
Seebeck coefficients are obtained as the ratio of voltage difference
and temperature difference when there are no electrical currents,
i.e., 
\begin{align*}
S_{j3} & =-\left(\frac{\delta\mu_{j}}{e\delta T}\right)_{J_{k}^{N}=0\;\forall k}.
\end{align*}
We find 
\begin{align}
S_{13} & =\frac{1}{eT}\frac{L_{1,3}L_{2,2}-L_{1,2}L_{2,3}}{L_{1,1}L_{2,2}-L_{1,2}L_{2,1}},\nonumber \\
S_{23} & =\frac{1}{eT}\frac{L_{1,1}L_{2,3}-L_{1,3}L_{1,2}}{L_{1,1}L_{2,2}-L_{1,2}L_{2,1}}.\label{eq:seebeck_three}
\end{align}
The Peltier coefficient\emph{ }is related to Seebeck coefficient
by $\Pi_{3j}=TS_{j3}.$ Thermal conductance is defined as the ratio
of heat current and temperature difference when the particle current
is zero, i.e. 
\begin{align}
K & =\left(\frac{J_{1}^{Q}}{\delta T}\right)_{\substack{J_{k}^{N}=0\;\forall k}
}=\frac{1}{T^{2}}\lrp{L_{3,3}-L_{3,1}S_{11}-L_{3,2}S_{21}}.\label{eq:heat_conduc_three}
\end{align}
From Eqs.$\;$(\ref{eq:ele_cond_three}-\ref{eq:heat_conduc_three}),
we recognize that the electrical conductance between the left reservoir
and the chain is still given by $G_{11}^{e}=L_{1,1}/T$, which is
identical to the two-terminal case. Nevertheless, the expressions
for Seebeck coefficients $S_{j3}$ and thermal conductance $K$ are
different from the two-terminal case$\;$\citep{blundell2010}.

\begin{figure}
\includegraphics[scale=0.28]{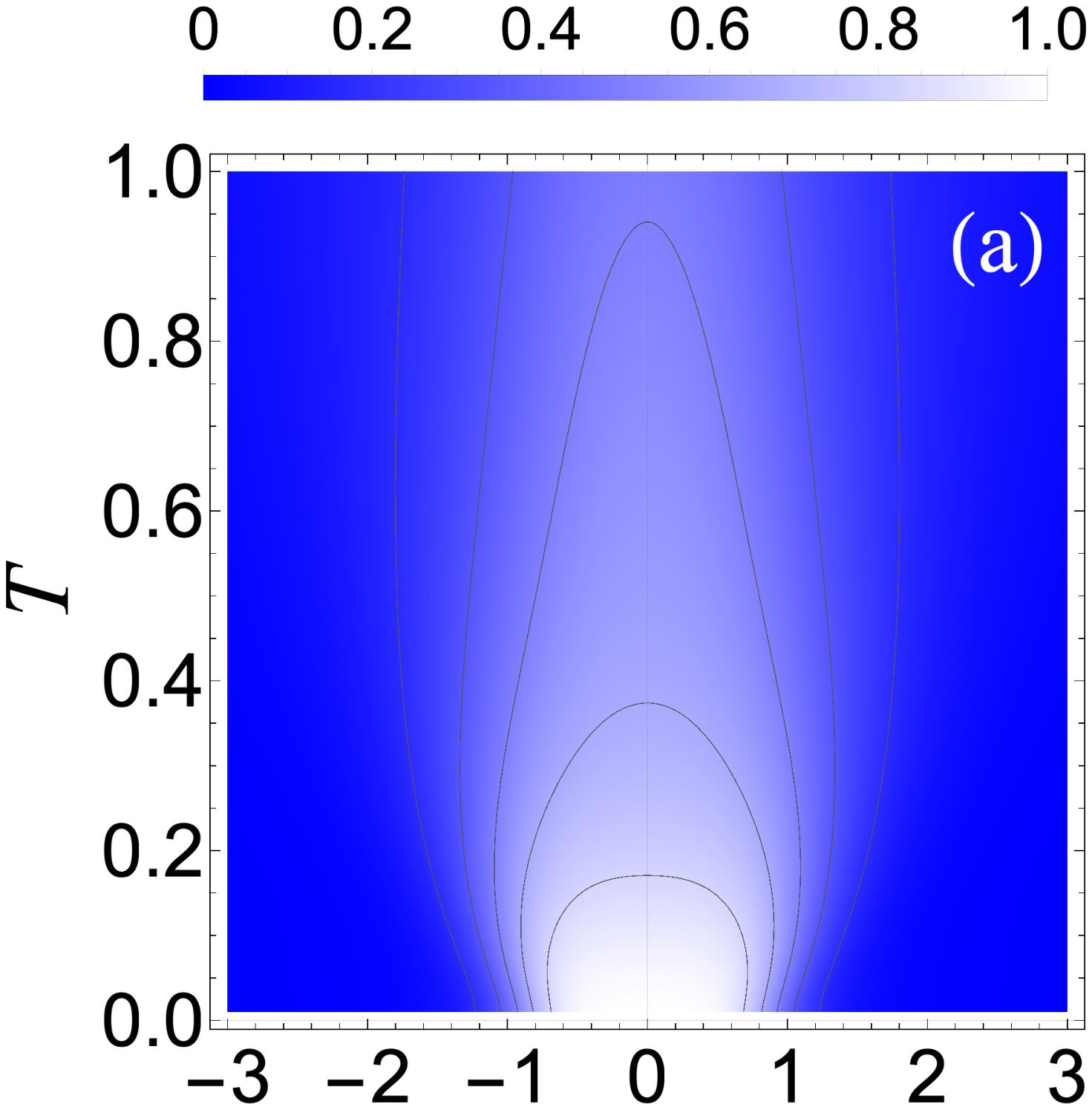}\includegraphics[scale=0.28]{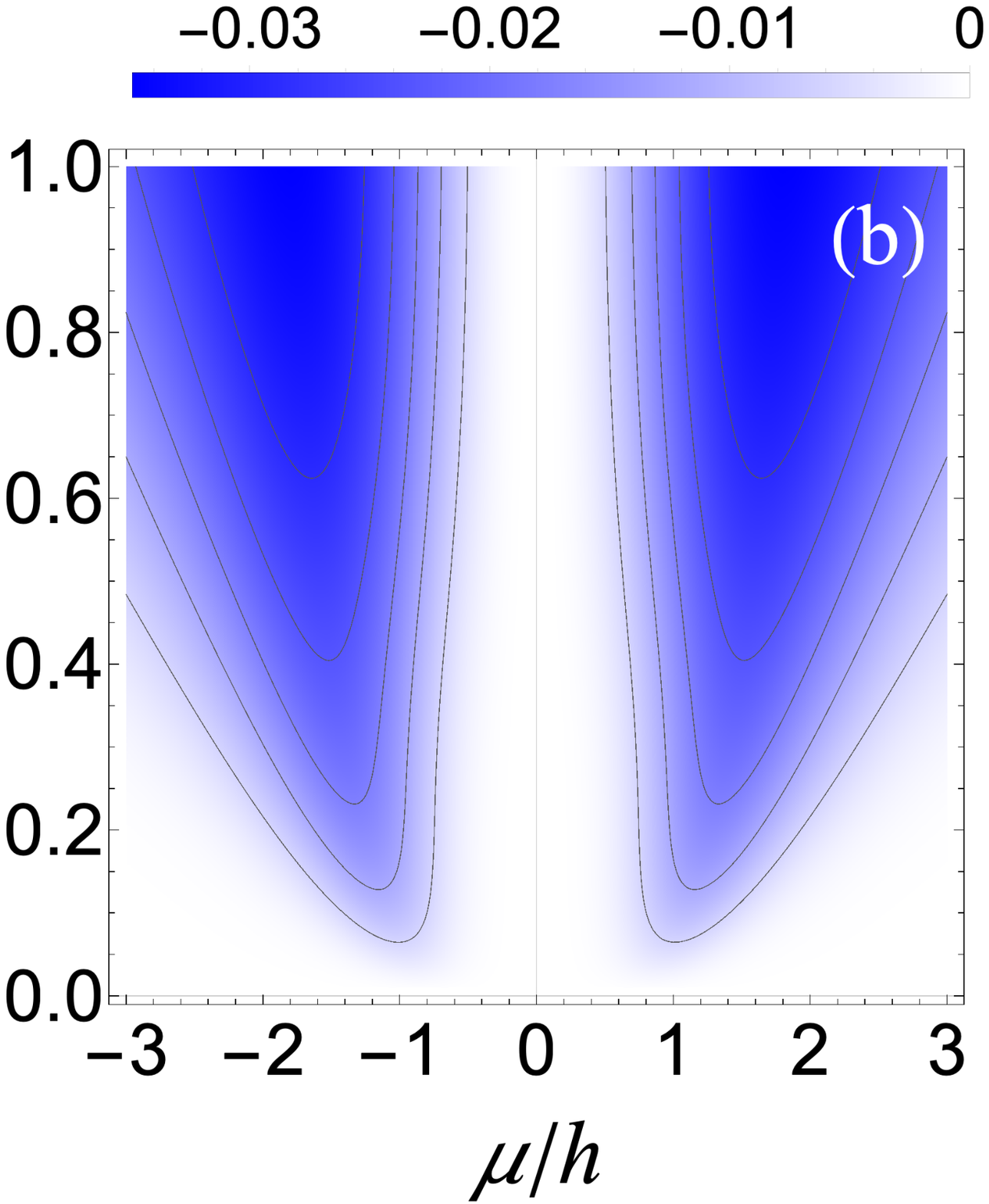}

\includegraphics[scale=0.28]{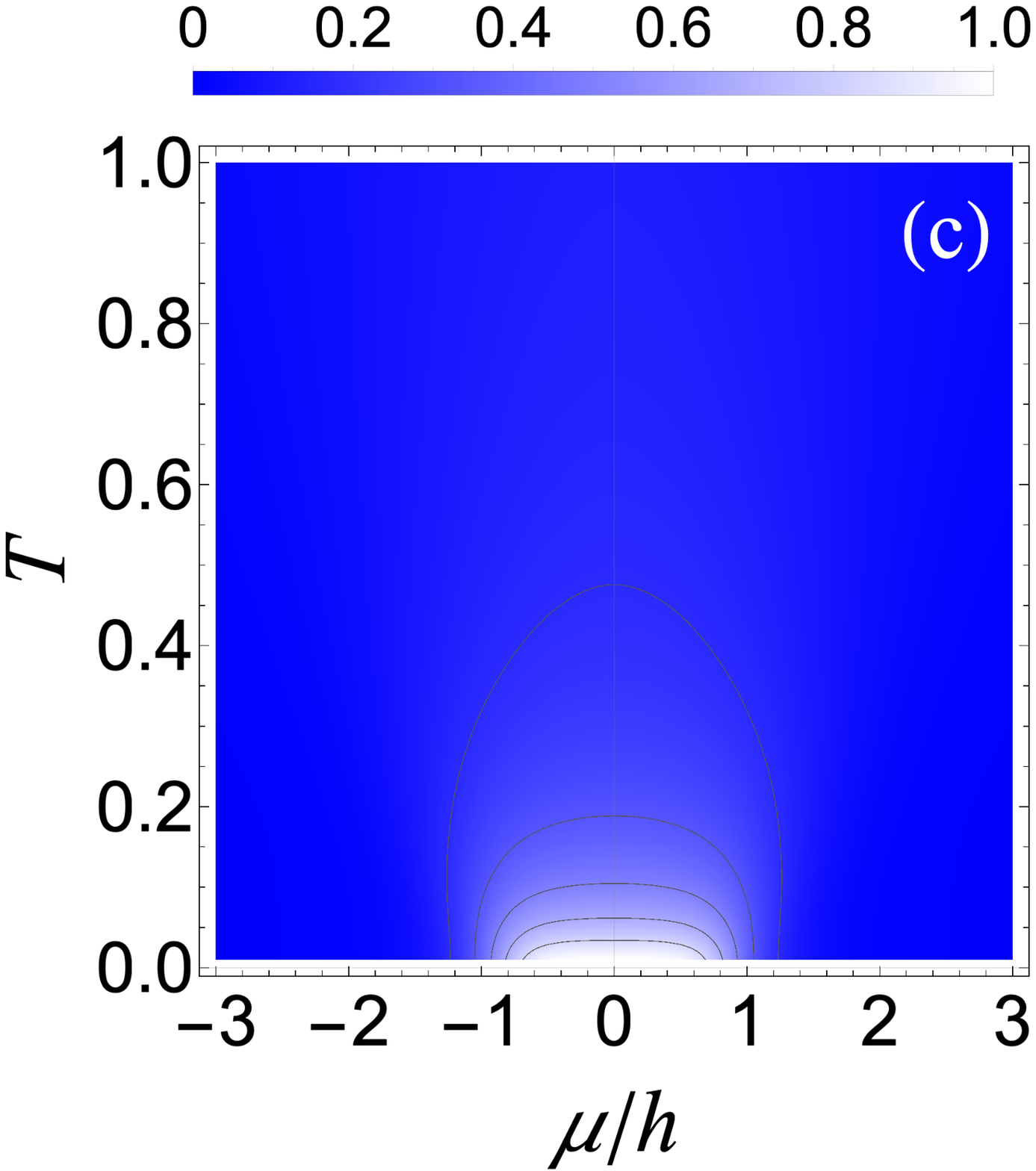}\includegraphics[scale=0.28]{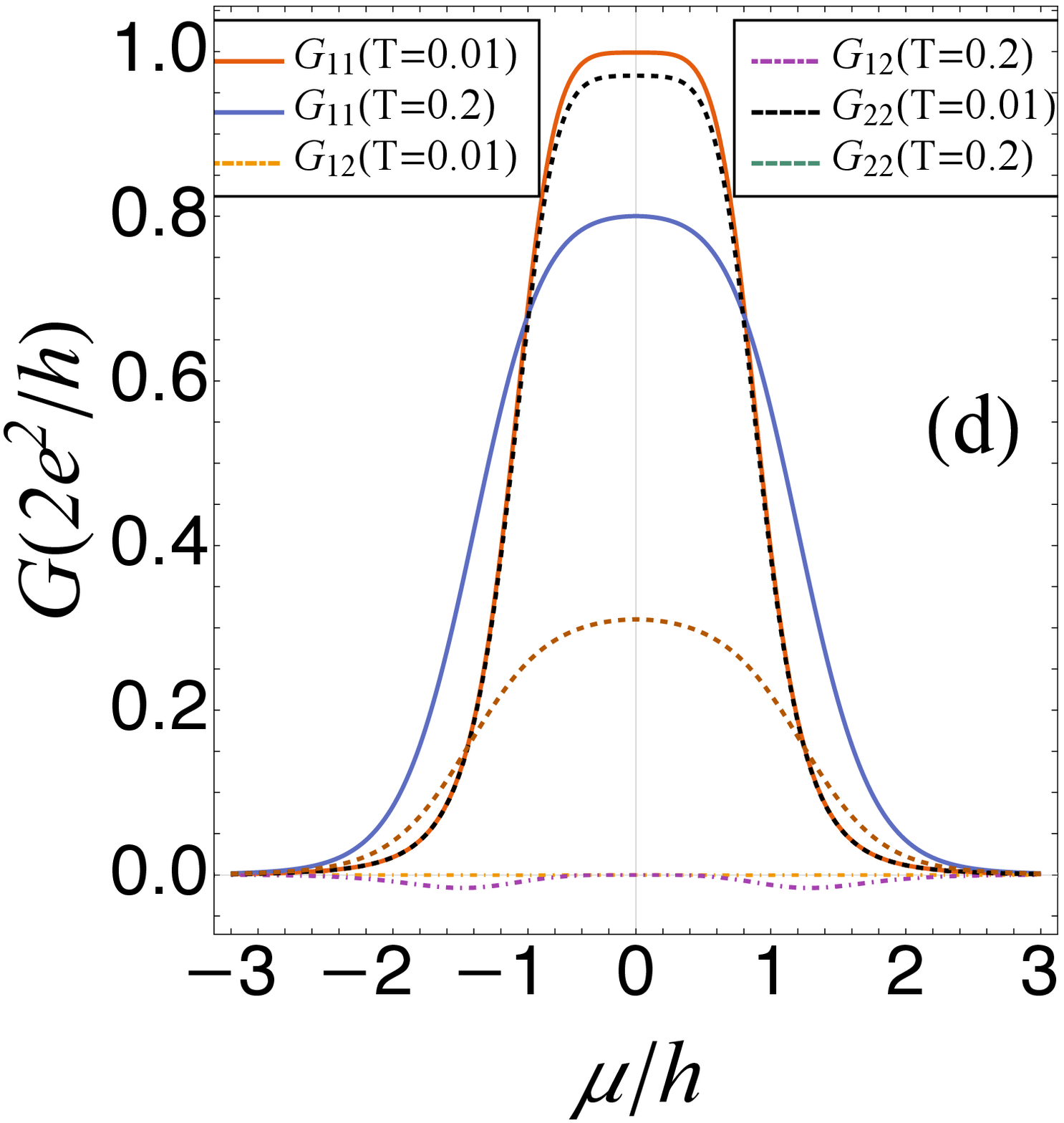}

\caption{The electrical conductance $G_{11}$, $G_{12}$ and $G_{22}$ for
a three-site Kitaev chain. The parameters are set to be $h=\Delta=1$,
and we adopt asymmetric coupling $\Gamma_{1}=0.5,\Gamma_{2}=0.1.$
 (a-c) are for $G_{11}$, $G_{12}$, and $G_{22}$ respectively.
(d) Cross section of electrical conductances for low temperature $T=0.01$
and high temperature $T=0.2$. The local conductances $G_{11}$ and
$G_{22}$ are quantized at low temperature when $|\mu|<h,$ which
indicates the current component of LAR dominates the transport process.
\label{fig:ele_conduc}}
\end{figure}

We fix the hopping amplitude $h=1$, and plot electrical conductance
$G$ for different $\mu$ at different temperatures for a three-site
model in Fig.$\;$\ref{fig:ele_conduc}. We adopt the asymmetric effective
coupling strength $\Gamma_{1}=0.5$, $\Gamma_{2}=0.1$$\;$\footnote{The relation between $\Gamma_{\alpha}$ and $\lambda_{\alpha j}$
can be found in our previous paper \citep{Zhang2021pre}}. We find the local electrical conductance$\;${[}Fig.$\;$\ref{fig:ele_conduc}(a,
c){]} is nearly unity for low $T$ in the region $|\mu|<h.$ It is
consistent with the spectrum of a three-site model in Fig.$\;$\ref{fig:The-energy-spectrum}
which shows that the nanowire hosts Majorana modes at two ends of
the wire when $|\mu|<h$. In contrast, the nonlocal electrical conductance$\;${[}Fig.$\;$\ref{fig:ele_conduc}(b){]}
is small at low temperature and is nonzero near the gap-opening region.
The low-temperature feature of the electrical conductances of the
left and right reservoirs is quite similar even though the coupling
strength are asymmetric. As the temperature increases, high-energy
modes begin to get involved, and the asymmetry in the coupling strength
affect the conductance dramatically. For example, at $T=0.2$, the
electrical conductance of the left reservoir $G_{11}$ is nearly three
times of the right one $G_{22}$$\;${[}see Fig.$\;$\ref{fig:ele_conduc}(d){]}.

\begin{figure}
\includegraphics[scale=0.28]{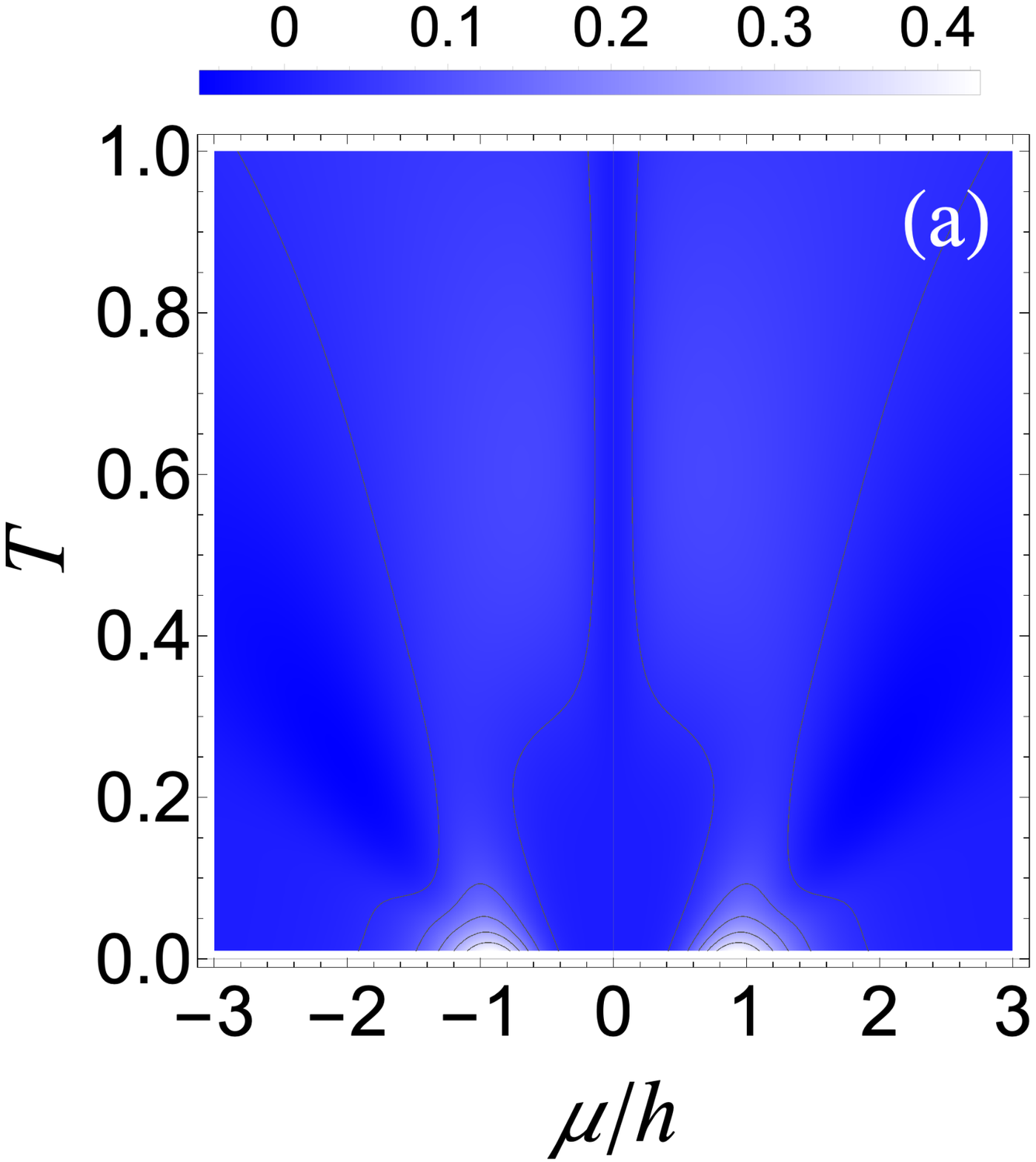}\includegraphics[scale=0.28]{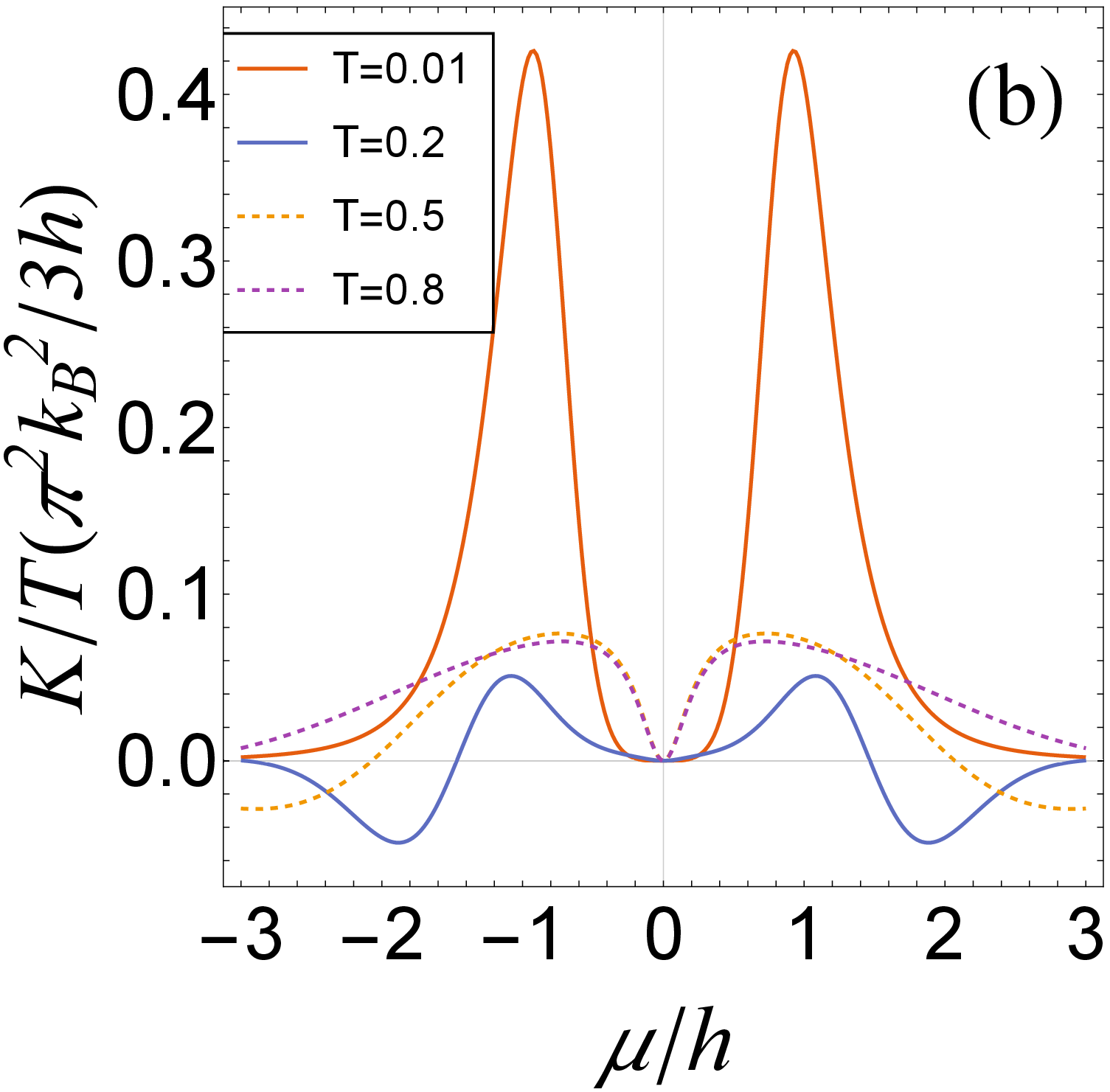}\caption{Thermal conductivity of a three-site Kitaev chain. The parameters
are $h=\Delta=1$, and we adopt asymmetric coupling $\Gamma_{1}=0.5,\Gamma_{1}=0.1$.
(a) The temperature and chemical potential dependence of $K/T$. \label{fig:Heat_conduc}}
\end{figure}

In Fig.$\;$\ref{fig:Heat_conduc}, we show the thermal conductance
$K/T$ as a function of $T$ and $\mu/h$. We see that the $K/T$
is nearly zero in the region $|\mu|<h$ and increases dramatically
near the gap opening region $|\mu|/h\sim1$, which is different from
the behavior of local electrical conductance. The zero value of thermal
conductance is due to the fact that LAR does not transfer energy.
It can also be seen in Fig.$\;$\ref{fig:Heat_conduc}(b). When $\mu=0$,
the system is in the exactly solvable case. Two Majorana modes are
perfectly localized at two ends of the chain. NT and CAR vanish and
only LAR is present. There is no energy transport, so the heat conductance
is always zero regardless of the temperature. Another interesting
feature is that the peak at low temperature $T=0.02$ is close to
half of a thermal conductance quantum $(1/2)\pi^{2}k_{B}^{2}/3h.$
Similar feature is also seen in the two MZMs case$\;$(see Appendix
\ref{sec:Two-Terminal-Majorana-Junction}). In the two MZMs case,
we prove that the half quantization is exact when $\epsilon_{M}/\Gamma=2$
for large gap $\epsilon_{M}$ and large coupling $\Gamma$$\;$(see
Appendix \ref{sec:Two-Terminal-Majorana-Junction}). Here in the Kitaev
chain, it is still an open question if the half quantization is exact$\;$(in
the thermodynamic limit) or rather accidental, and if it is a feature
of Majorana physics or anything else. 

\begin{figure}
\includegraphics[scale=0.28]{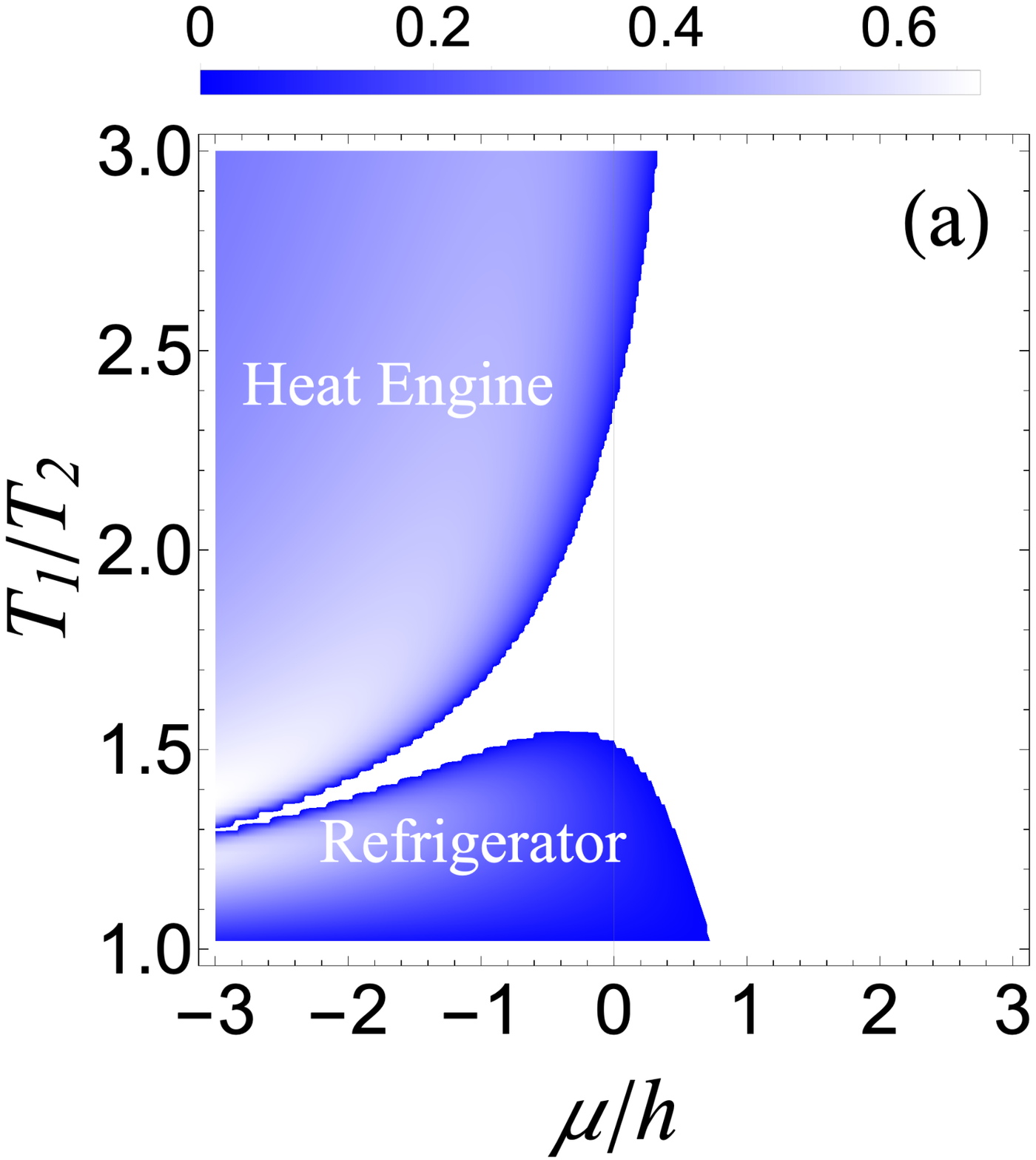}\includegraphics[scale=0.28]{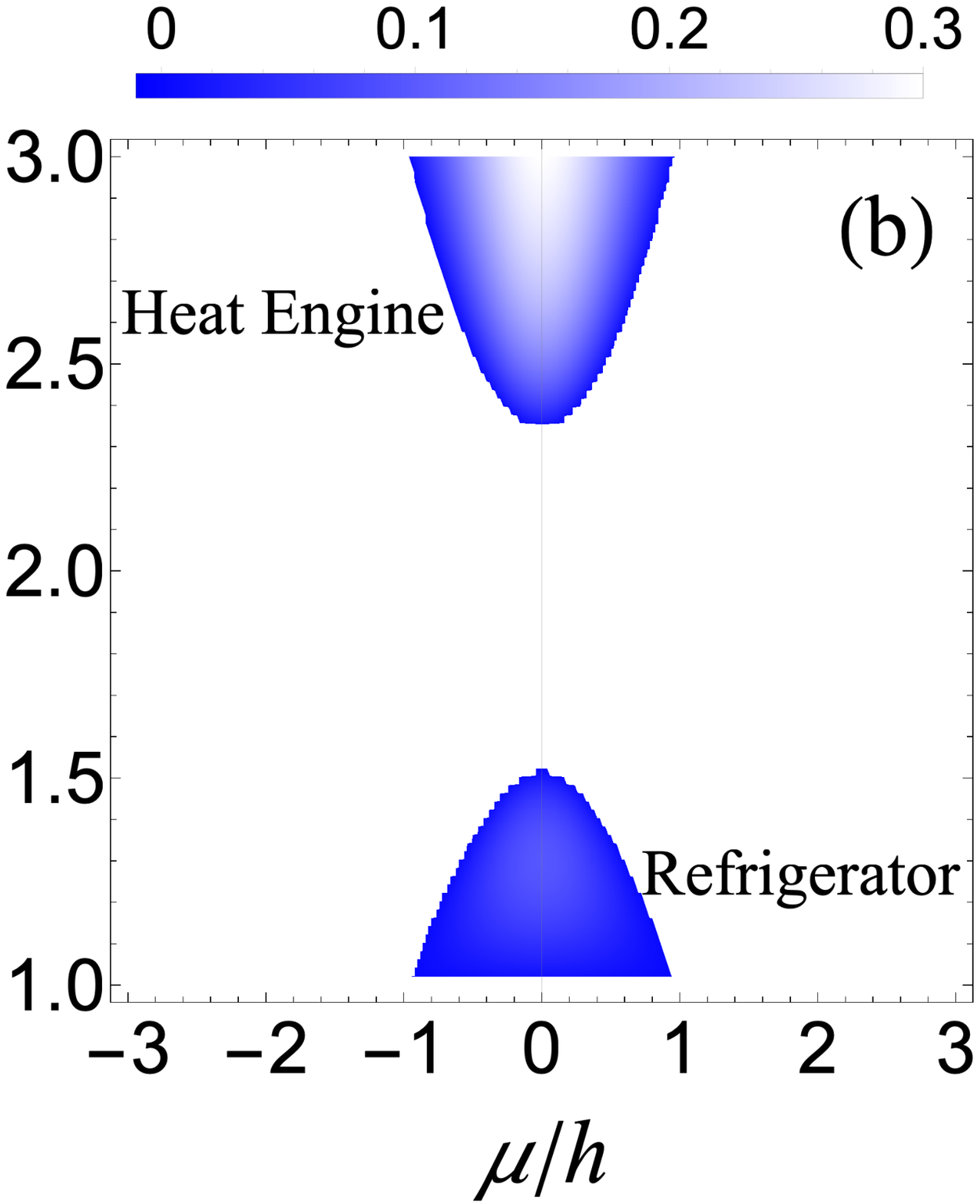}\caption{Efficiency of a four-site Kitaev chain as a heat engine or refrigerator.
The left reservoir is always hotter than the right reservoir. The
system operates as a heat engine when $J_{1}^{Q}>0$ and $\dot{W}>0$
and as a refrigerator when $J_{2}^{Q}>0$ and $\dot{W}<0$. (a) Only
NT is present$\;$($h=1$, $\Delta=0$, $\mu_{2}=1$). (b) Only CAR
is present$\;$($h=0$, $\Delta=1$, $\mu_{2}=-1$). The efficiency
is symmetric about $\mu=0$ in (b). The other parameters are chosen
to be $\Gamma_{1}=\Gamma_{2}=0.5$, $T_{2}=1,$ $\mu_{1}=2$. All
efficiencies are measured in the corresponding Carnot efficiencies,
i.e., $\eta_{{\rm {\rm C},HE}}=1-T_{2}/T_{1}$ and $\eta_{{\rm C},{\rm Ref}}=1/(T_{1}/T_{2}-1).$\label{fig:eff_4sites}}
\end{figure}

As a thermoelectric device, we can discuss its power and efficiency.
According to the first law of thermodynamics, the work power $\dot{W}$
is defined as$\;$(positive work means that the system outputs power)
\begin{align}
-\dot{W}+\sum_{\alpha=1,2}J_{\alpha}^{Q} & =0\nonumber \\
\implies\dot{W}=\sum_{\alpha=1,2}J_{\alpha}^{Q} & =-\sum_{\alpha=1,2}\mu_{\alpha}J_{\alpha}^{N},\label{eq:work_power_def}
\end{align}
where $J_{2}^{Q}\equiv J_{2}^{E}-\mu_{2}J_{2}^{N}$ and we use the
energy conservation $J_{1}^{E}+J_{2}^{E}=0$ in the second equality
of Eq.$\;$(\ref{eq:work_power_def}). Following Ref.$\;$\citep{Mazza2014NJP},
the efficiency of a three-terminal system operating as a heat engine
is defined as the ratio between the work power and the positive heat
currents
\[
\eta_{{\rm HE}}=\frac{\dot{W}}{\sum_{\alpha+}J_{\alpha+}^{Q}},
\]
where $\alpha+$ denotes a positive current. And if it operates as
a refrigerator, the efficiency is defined as the ratio between the
positive heat current and the work power supplied to the system
\[
\eta_{{\rm Ref}}=\frac{\sum_{\alpha+}J_{\alpha+}^{Q}}{-\dot{W}}.
\]
 The corresponding Carnot efficiencies are $\eta_{{\rm {\rm C},HE}}=1-T_{2}/T_{1}$
and $\eta_{{\rm C},{\rm Ref}}=1/(T_{1}/T_{2}-1)$ for the heat engine
and the refrigerator, respectively. As in Sec.$\;$\ref{sec:Fluctuation-Relation},
we discuss three special cases, in which only one current component
is present. We assume that $\mu_{1}>\mu_{2}$ and $T_{1}<T_{2}$ without
loss of generality. In the first$\;$(NT) case, the two particle currents
are opposite to each other $J_{1}^{N}=-J_{2}^{N}.$ The work power
is $\dot{W}=-(\mu_{1}-\mu_{2})J_{1}^{N}=-\delta\mu J_{1}^{N}$ with
$\delta\mu=\mu_{1}-\mu_{2}.$ In order to generate a positive work,
the signs of $J_{1}^{N}$ and $\delta\mu_{1}$ must be opposite to
each other. From the FR$\;$(\ref{eq:FR_NT}), a negative $J_{1}^{N}$
requires $\beta_{1}\mu_{1}<\beta_{2}\mu_{2}$ which implies that $\mu_{2}>0$
and $\beta_{2}>\beta_{1}\mu_{1}/\mu_{2}$. In the second$\;$(CAR)
case, $J_{1}^{N}=J_{2}^{N}.$ The work power is $\dot{W}=-(\mu_{1}+\mu_{2})J_{1}^{N}.$
From the FR (\ref{eq:FR_CAR}), a negative $J_{1}^{N}$ requires $\beta_{1}\mu_{1}+\beta_{2}\mu_{2}<0,$
which implies $\mu_{2}<0$, $\beta_{2}>\beta_{1}\mu_{1}/|\mu_{2}|$
if we let $\mu_{1}>0.$ In the third$\;$(LAR) case, $J_{1}^{E}=J_{2}^{E}=0$.
The work is $\dot{W}=-\mu_{1}J_{1}^{N}-\mu_{2}J_{2}^{N}.$ From FR$\;$(\ref{eq:FR_MZM}),
$\beta_{\alpha}\mu_{\alpha}$ has the same sign as current $J_{\alpha}^{N}$.
Hence, the work power is always negative and the Kitaev chain can't
serve as a useful heat engine. In Fig.$\;$\ref{fig:eff_4sites},
we fix $\beta_{2}=1,$ $\mu_{1}=2$, and $\mu_{2}=\pm1$$\;${[}plus
for Fig.$\;$\ref{fig:eff_4sites}(a); minus for Fig.$\;$\ref{fig:eff_4sites}(b){]},
then vary $\beta_{2}/\beta_{1}=T_{1}/T_{2}$ and $\mu.$ We see that
the efficiency $\eta$ is highly asymmetric about $\mu$ in the NT
case$\;${[}Fig.$\;$\ref{fig:eff_4sites}(a){]}, while it is symmetric
about $\mu$ in the CAR case. We further note that the maximum of
$\eta$ locates at $\mu=0$ in the CAR case. 

In the linear response regime, the power can be expressed as 
\[
\dot{W}=-\sum_{\alpha}\mu_{\alpha}J_{\alpha}^{N}=-\sum_{i,j=1,2}\mu_{i}\frac{L_{i,j}}{T}\mu_{j}-\mu_{i}L_{i,3}\frac{\delta T}{T^{2}}.
\]
Optimizing $\dot{W}$ at a fixed temperature gradient, we find the
maximum work power as 
\begin{equation}
\dot{W}_{{\rm max}}=\frac{1}{4}\delta T^{2}\left(G_{11}S_{11}^{2}+G_{22}S_{21}^{2}+2G_{12}S_{11}S_{21}\right),\label{eq:max_work_power}
\end{equation}
 with the condition
\[
\begin{pmatrix}\mu_{1}, & \mu_{2}\end{pmatrix}=eT\begin{pmatrix}S_{11}, & S_{21}\end{pmatrix}.
\]
Eq.$\;$(\ref{eq:max_work_power}) can be seen as a generalization
of the maximum power $\dot{W}'_{{\rm max}}=GS^{2}\delta T^{2}/4$
in two terminal case$\;$\citep{Whitney2018book}.
\begin{widetext}
Now we discuss the diffusivity. Here we only consider the diffusivity
of energy and particle transport of the left reservoir. The diffusivity
is symmetric to its index and has six independent components. Near
equilibrium, the diffusivities are 
\[
D_{jk}(0)=\intw f(\omega)[1-f(\omega)]\begin{pmatrix}\tdbbT_{1}+\tdbbT_{2}+2\tdbbT_{3} & -\tdbbT_{1}+\tdbbT_{2} & \omega(\tdbbT_{1}+\tdbbT_{2})\\
-\tdbbT_{1}+\tdbbT_{2} & \tdbbT_{1}+\tdbbT_{2}+2\tdbbT_{4} & \omega(-\tdbbT_{1}+\tdbbT_{2})\\
\omega(\tdbbT_{1}+\tdbbT_{2}) & \omega(-\tdbbT_{1}+\tdbbT_{2}) & \omega^{2}(\tdbbT_{1}+\tdbbT_{2})
\end{pmatrix}.
\]
They are equal to the linear response coefficients $L_{j,k}$$\;${[}Eq.$\;$(\ref{eq:linear_coeff_L11}-\ref{eq:linear_coeff_L33}){]}.
Thus, we explicitly verify the fluctuation-dissipation relation in
our model. In relevance to the experiment, the diffusivity is related
to the zero-frequency noise power by
\[
D(\bfA)=\half\mcS(0,V),
\]
where the zero-frequency noise power is defined as the Fourier transform
of the symmetric current correlation 
\[
\mcS_{ij}(\omega,V)\equiv\int dte^{i\omega t}\avg{(\hat{J}_{i}(t)-J_{i})(\hat{J}_{j}(0)-J_{j})+(\hat{J}_{j}(0)-J_{j})(\hat{J}_{i}(t)-J_{i})}.
\]
Here $\hat{J}_{j}=\partial_{t}\hat{X}_{j}$ is the current operator.
A quantized electrical conductance $G=2e^{2}/h=1/\pi$ (in natural
unit) implies a quantized $\mcS(0,V)/T=4/h=2/\pi$ by the fluctuation-dissipation
relation.
\end{widetext}

\subsection{Nonlinear response in Kitaev chain}

In this section, we go beyond the linear response regime, and check
the nonlinear response relation Eq.$\;$(\ref{eq:sec_order_response_relation})
explicitly.
\begin{widetext}
The second-order response coefficients $M_{i,jk}$ ($i,j,k=1,2,3$)
at zero affinity are given by 
\begin{align*}
M_{1,ij} & =\intw p_{0}(1-p_{0})(1-2p_{0})\begin{pmatrix}\tdbbT_{1}+\tdbbT_{2} & 0 & \omega(\tdbbT_{1}+\tdbbT_{2}+2\tdbbT_{3})\\
0 & -(\tdbbT_{1}+\tdbbT_{2}) & 0\\
\omega(\tdbbT_{1}+\tdbbT_{2}+2\tdbbT_{3}) & 0 & \omega^{2}(\tdbbT_{1}+\tdbbT_{2})
\end{pmatrix},\\
M_{2,ij} & =\intw p_{0}(1-p_{0})(1-2p_{0})\begin{pmatrix}-\tdbbT_{1}+\tdbbT_{2} & 0 & 0\\
0 & \tdbbT_{1}-\tdbbT_{2} & \omega(-\tdbbT_{1}+\tdbbT_{2})\\
0 & \omega(-\tdbbT_{1}+\tdbbT_{2}) & \omega^{2}(-\tdbbT_{1}+\tdbbT_{2})
\end{pmatrix},\\
M_{3,ij} & =\intw p_{0}(1-p_{0})(1-2p_{0})\begin{pmatrix}\omega(\tdbbT_{1}+\tdbbT_{2}) & 0 & \omega^{2}(\tdbbT_{1}+\tdbbT_{2})\\
0 & -\omega(\tdbbT_{1}+\tdbbT_{2}) & 0\\
\omega^{2}(\tdbbT_{1}+\tdbbT_{2}) & 0 & \omega^{3}(\tdbbT_{1}+\tdbbT_{2})
\end{pmatrix}.
\end{align*}
 Accordingly, the derivative of diffusivities $D_{ij}$ are given
by 
\begin{align*}
\frac{\partial D_{1i}}{\partial A_{j}} & =\half\intw p_{0}(1-p_{0})(1-2p_{0})\begin{pmatrix}\tdbbT_{1}+\tdbbT_{2} & \tdbbT_{1}-\tdbbT_{2} & \omega(\tdbbT_{1}+\tdbbT_{2}+4\tdbbT_{3})\\
-\tdbbT_{1}+\tdbbT_{2} & -\tdbbT_{1}-\tdbbT_{2} & \omega(-\tdbbT_{1}+\tdbbT_{2})\\
\omega(\tdbbT_{1}+\tdbbT_{2}) & \omega(\tdbbT_{1}-\tdbbT_{2}) & \omega^{2}(\tdbbT_{1}+\tdbbT_{2})
\end{pmatrix},\\
\frac{\partial D_{2i}}{\partial A_{j}} & =\half\intw p_{0}(1-p_{0})(1-2p_{0})\begin{pmatrix}-\tdbbT_{1}+\tdbbT_{2} & -\tdbbT_{1}-\tdbbT_{2} & \omega(-\tdbbT_{1}+\tdbbT_{2})\\
\tdbbT_{1}+\tdbbT_{2} & \tdbbT_{1}-\tdbbT_{2} & \omega(\tdbbT_{1}+\tdbbT_{2}+4\tdbbT_{4})\\
\omega(-\tdbbT_{1}+\tdbbT_{2}) & -\omega(\tdbbT_{1}+\tdbbT_{2}) & -\omega^{2}(\tdbbT_{1}-\tdbbT_{2})
\end{pmatrix},\\
\frac{\partial D_{3i}}{\partial A_{j}} & =\half\intw p_{0}(1-p_{0})(1-2p_{0})\begin{pmatrix}\omega(\tdbbT_{1}+\tdbbT_{2}) & \omega(\tdbbT_{1}-\tdbbT_{2}) & \omega^{2}(\tdbbT_{1}+\tdbbT_{2})\\
\omega(-\tdbbT_{1}+\tdbbT_{2}) & -\omega(\tdbbT_{1}+\tdbbT_{2}) & -\omega^{2}(\tdbbT_{1}-\tdbbT_{2})\\
\omega^{2}(\tdbbT_{1}+\tdbbT_{2}) & \omega^{2}(\tdbbT_{1}-\tdbbT_{2}) & \omega^{3}(\tdbbT_{1}+\tdbbT_{2})
\end{pmatrix}.
\end{align*}
 It is easy to see that the nonlinear response relation Eq.$\;$(\ref{eq:sec_order_response_relation})
is satisfied, namely
\[
M_{i,jk}=\frac{\partial D_{ij}}{\partial A_{k}}+\frac{\partial D_{3k}}{\partial A_{j}}.
\]
Higher-order response relations can be checked similarly.
\end{widetext}

\section{Summary}

In this article, we analysize transport in 1D open Kitaev chain. We
obtain a general form of MGF of energy and particle transport at finite
temperature. The explicit expression of MGF allows us to extract the
fluctuation relations in a straightforward manner. The energy current
is carried by the particles involved in the NT process and the CAR
process, while the particle current is also carried by the LAR process
in addition to the above two processes. We find that the joint distribution
of particle and energy currents obeys different fluctuation relations
in different regions of the parameter space as a result of $U$(1)
symmetry breaking and energy conservation. Moreover, we study the
response properties of the Kitaev chain. Explicitly, we calculate
the response coefficients, and find that they are consistent with
the relations derived from the fluctuation relation. In addition,
in the linear response regime, we treat the open Kitaev chain as a
three-terminal system instead of two-terminal system and discuss its
thermoelectrical properties. The electrical conductance is quantized
when the Kitaev chain hosts two Majorana modes at two ends as expected.
The thermal conductance, however, exhibits a peak$\;$(up to half
thermal conductance quantum) around the region of the gap opening.
The work power and the operation of thermoelectric device$\;$(based
on the Kitaev chain) is also discussed. We find that a Kitaev chain
in the topological superconductor phase always consumes energy, but
it can operate as a heat engine or refrigerator otherwise. %
\begin{comment}
As a result, the only nonzero response coefficient of LAR is the linear
response coefficient with respect to the electrical voltage.
\end{comment}

\begin{comment}
We pick up the MGF of LAR 
\[
Z_{{\rm LAR,1}}(\omega)=T_{43}\lrb{n_{1e}\barn_{1h}\lrp{e^{2i\xi_{1}}-1}+\barn_{1e}n_{1h}\lrp{e^{-2i\xi_{1}}-1}}.
\]
The process is equivalent to a binomial process.
\end{comment}
\begin{comment}
For a binomial process with probability $p$, the $n$-th cumulant
$\kappa_{n}$ can be obtained by recursion relation 
\[
\kappa_{n}=\kappa_{n-1}\frac{\partial\kappa_{n}}{\partial p}.
\]
For example,
\end{comment}

\begin{comment}
The linear transport relation is$\;$(concept in thermal physicsBook)
\begin{align*}
J_{e} & =\mcL_{NN}V+\mcL_{NE}\nabla T,\\
J_{Q} & =\mcL_{EN}V+\mcL_{EE}\nabla T.
\end{align*}
 In our case, $-eV=\mu_{1}-\mu_{2}.$
\end{comment}

\begin{acknowledgments}
We acknowledges support from the National Science Foundation of China
under grants 11775001, 11825501, and 12147162.
\end{acknowledgments}

\appendix

\section{Two-Terminal Majorana Junction\label{sec:Two-Terminal-Majorana-Junction}}

In this appendix, we study the transport of two Majorana modes localized
at two ends of a nanowire. This model has been extensively studied
in the literature, since it is simple enough but still captures the
main features of the Majorana physics. The Hamiltonian of the whole
system is composed of three parts $\hatH=\hatH_{M}+\sum_{\alpha=1,2}\hatH_{\alpha}+\hatH_{I}',$
where $\hatH_{\alpha}$ is given by Eq.$\;$(\ref{eq:Ham_of_reservoirs})
and 
\begin{align*}
\hatH_{M} & =\frac{i}{2}\epsilon_{M}\hat{\gamma}_{1}\hat{\gamma}_{2},\\
\hatH_{I}' & =\sum_{j}\left(t_{1j,1}\coc_{1j}\hat{\gamma}_{1}+t_{2j,2}\coc_{2j}\hat{\gamma}_{2}+{\rm H.c.}\right).
\end{align*}
Here, $\epsilon_{M}$ is the energy gap of the MZMs, and $\hatH_{I}'$
describes the coupling between the reservoirs and the nearest Majorana
mode $\hat{\gamma}_{\alpha}.$ The Majorana modes satisfy the anticommutation
relation $\lrc{\aogam_{\alpha},\aogam_{\alpha'}}=2\delta_{\alpha\alpha'},$
and can be combined to a Dirac fermion $\cod=(\hat{\gamma}_{1}+i\hat{\gamma}_{2})/2$,
$\aod=(\aogam_{1}-i\aogam_{2})/2$ which satisfies $\lrc{\aod,\cod}=1.$
\begin{widetext}
We use the Keldysh functional integral and obtain the MGF of this
system. The MGF takes the same form as Eq.$\;$(\ref{eq:MGF}) but
the expressions of the components are slightly different 
\begin{align}
Z_{M,\rmNT}(\xi_{1}-\xi_{2}) & =\bbC_{1}+\bbT_{1}\left[n_{1e}\barn_{2e}(e^{i(\xi_{1}-\xi_{2})}e^{i\omega\eta}-1)+\barn_{1e}n_{2e}(e^{-i(\xi_{1}-\xi_{2})}e^{-i\omega\eta}-1)\right]\nonumber \\
 & \quad\quad\;+\bar{\bbT}_{1}\left[n_{1h}\barn_{2h}(e^{-i(\xi_{1}-\xi_{2})}e^{i\omega\eta}-1)+\barn_{1h}n_{2h}(e^{i(\xi_{1}-\xi_{2})}e^{-i\omega\eta}-1)\right],\label{eq:MGF_twoMZM_NT}\\
Z_{M,\rmCAR}(\xi_{1}+\xi_{2}) & =\bbC_{2}+\bbT_{2}\left[n_{1e}\barn_{2h}(e^{i(\xi_{1}+\xi_{2})}e^{i\omega\eta}-1)+\barn_{1e}n_{2h}e^{-i(\xi_{1}+\xi_{2})}e^{-i\omega\eta}-1)\right]\nonumber \\
 & \quad\quad\;+\bar{\bbT}_{2}\left[n_{1h}\barn_{2e}(e^{-i(\xi_{1}+\xi_{2})}e^{i\omega\eta}-1)+\barn_{1h}n_{2e}(e^{i(\xi_{1}+\xi_{2})}e^{-i\omega\eta}-1)\right],\label{eq:MGF_twoMZM_CAR}\\
Z_{M,\rmLAR}(\xi_{1},\xi_{2}) & =\left\{ \bbC_{3}+\bbT_{3}\left[n_{1e}\barn_{1h}(e^{2i\xi_{1}}-1)+\barn_{1e}n_{1h}(e^{-2i\xi_{1}}-1)\right]\right\} \nonumber \\
 & \quad\times\left\{ \bbC_{4}+\bbT_{4}\left[n_{2e}\barn_{2h}(e^{2i\xi_{2}}-1)+\barn_{2e}n_{2h}(e^{-2i\xi_{2}}-1)\right]\right\} .\label{eq:MGF_twoMZM_LAR}
\end{align}
The reflection and transmission coefficients are given by
\begin{align*}
\bbT_{1} & =\bar{\bbT}_{1}=\bbT_{2}=\bar{\bbT}_{2}=4\Gamma_{1}\Gamma_{2}\epsilon_{M}^{2},\quad\bbT_{3}=4\Gamma_{1}^{2},\quad\bbT_{4}=4\Gamma_{2}^{2},\\
\bbC_{3} & =4\Gamma_{1}^{2}+\omega^{2},\quad\bbC_{4}=4\Gamma_{2}^{2}+\omega^{2},\quad\bbC_{1}+\bbC_{2}=\epsilon_{M}^{2}(8\Gamma_{1}\Gamma_{2}+\epsilon_{M}^{2}-2\omega^{2}).
\end{align*}
We assume the effective coupling strengths are equal $\Gamma_{1}=\Gamma_{2}=\Gamma$.
The particle currents from the left and right reservoirs are 
\begin{align}
J_{1}^{N} & =\intw\frac{4\Gamma^{2}(n_{1e}\barn_{1h}-n_{1h}\barn_{1e})\left(4\Gamma^{2}+\epsilon_{M}^{2}+\omega^{2}\right)}{\left(4\Gamma^{2}+\omega^{2}\right)^{2}+\left(8\Gamma^{2}-2\omega^{2}\right)\epsilon_{M}^{2}+\epsilon_{M}^{4}}=\intw\;\bbT_{N}(\omega)(n_{1e}-n_{1h}),\label{eq:current_1N_MZM}\\
J_{2}^{N} & =\intw\frac{4\Gamma^{2}(n_{2e}\barn_{2h}-n_{2h}\barn_{2e})\left(4\Gamma^{2}+\epsilon_{M}^{2}+\omega^{2}\right)}{\left(4\Gamma^{2}+\omega^{2}\right)^{2}+\left(8\Gamma^{2}-2\omega^{2}\right)\epsilon_{M}^{2}+\epsilon_{M}^{4}}=\intw\;\bbT_{N}(\omega)(n_{2e}-n_{2h}),\label{eq:current_2N_MZM}
\end{align}
where the transmission coefficient is 
\[
\bbT_{N}(\omega)=\frac{4\Gamma^{2}\left(4\Gamma^{2}+\epsilon_{M}^{2}+\omega^{2}\right)}{\left(4\Gamma^{2}+\omega^{2}\right)^{2}+\left(8\Gamma^{2}-2\omega^{2}\right)\epsilon_{M}^{2}+\epsilon_{M}^{4}}.
\]
The energy currents from the left and the right reservoirs are 
\begin{align*}
J_{1}^{E} & =\intw\omega\frac{4\Gamma^{2}\omega\epsilon_{M}^{2}(n_{1e}+n_{1h}-n_{2e}-n_{2h})}{\left(4\Gamma^{2}+\omega^{2}\right)^{2}+\left(8\Gamma^{2}-2\omega^{2}\right)\epsilon_{M}^{2}+\epsilon_{M}^{4}}=\intw\;2\omega\bbT_{E}(\omega)(n_{1e}-n_{2e}),\\
J_{2}^{E} & =\intw\frac{4\Gamma^{2}\omega\epsilon_{M}^{2}(-n_{1e}-n_{1h}+n_{2e}+n_{2h})}{\left(4\Gamma^{2}+\omega^{2}\right)^{2}+\left(8\Gamma^{2}-2\omega^{2}\right)\epsilon_{M}^{2}+\epsilon_{M}^{4}}=\intw\;2\omega\bbT_{E}(\omega)(n_{2e}-n_{1e})
\end{align*}
with 
\[
\bbT_{E}(\omega)=\frac{4\Gamma^{2}\epsilon_{M}^{2}}{\left(4\Gamma^{2}+\omega^{2}\right)^{2}+\left(8\Gamma^{2}-2\omega^{2}\right)\epsilon_{M}^{2}+\epsilon_{M}^{4}}.
\]
The net effect of the NT and the CAR in the particle current $J_{\alpha}^{N}$
is to convert an electron in the left reservoir to an hole in the
same reservoir. It can be seen by considering current $j(\omega)$
through a single channel $\omega$. From the MGF Eqs.$\;$(\ref{eq:MGF_twoMZM_NT},\ref{eq:MGF_twoMZM_CAR}),
the current components for a single channel $\omega$ from NT and
CAR are 
\[
j_{\rmNT}(\omega)=\bbT_{1}(n_{1e}-n_{2e})-\bar{\bbT}_{1}(n_{1h}-n_{2h}),\quad j_{\rmCAR}(\omega)=\bbT_{2}(n_{1e}-n_{2h})-\bar{\bbT}_{2}(n_{1h}-n_{2e}).
\]
Since $\bbT_{1}=\bbT_{2}=\bar{\bbT}_{1}=\bar{\bbT}_{2},$ we have
$j_{\rmNT}+j_{\rmCAR}=2\bbT_{1}(n_{1e}-n_{1h})$ which indicates the
whole process is equivalent to a LAR. Again, we see that $J_{1}^{N}\neq J_{2}^{N}$
generally and $J_{1}^{E}=-J_{2}^{E}.$ Previously, some studies, e.g.,
Refs.$\;$\citep{Lopez2014PRB,Ramos-Andrade2016PRB} treat the system
as a two-terminal system and use the Landauer-B\textipa{\" u}ttiker
formula, which is incorrect$\;$(in fact $G$ will be half of the
correct value) according to the full-counting statistics.
\end{widetext}

\subsection{Linear response regime}

In the framework of three-terminal system, the linear response matrix
for two MZMs reads 
\[
L=\begin{pmatrix}L_{1,1} & 0 & 0\\
0 & L_{2,2} & 0\\
0 & 0 & L_{3,3}
\end{pmatrix}
\]
with 
\begin{align}
L_{1,1} & =L_{2,2}=\intw\;\bbT_{N}(\omega)\frac{1}{2\cosh^{2}\frac{\beta_{2}\omega}{2}},\nonumber \\
L_{3,3} & =\intw\;2\omega^{2}\bbT_{E}(\omega)\frac{1}{4\cosh^{2}\frac{\beta_{2}\omega}{2}}.\label{eq:L33_TwoMZMs}
\end{align}
The electrical conductance is 
\[
G=\frac{e^{2}}{T}\begin{pmatrix}L_{1,1} & 0\\
0 & L_{2,2}
\end{pmatrix},
\]
which shows that there is no non-local conductance. The Seebeck coefficients
are all zero. Thermal conductance is given by 
\[
K=\frac{1}{T^{2}}L_{3,3}.
\]

\begin{figure}
\includegraphics[scale=0.28]{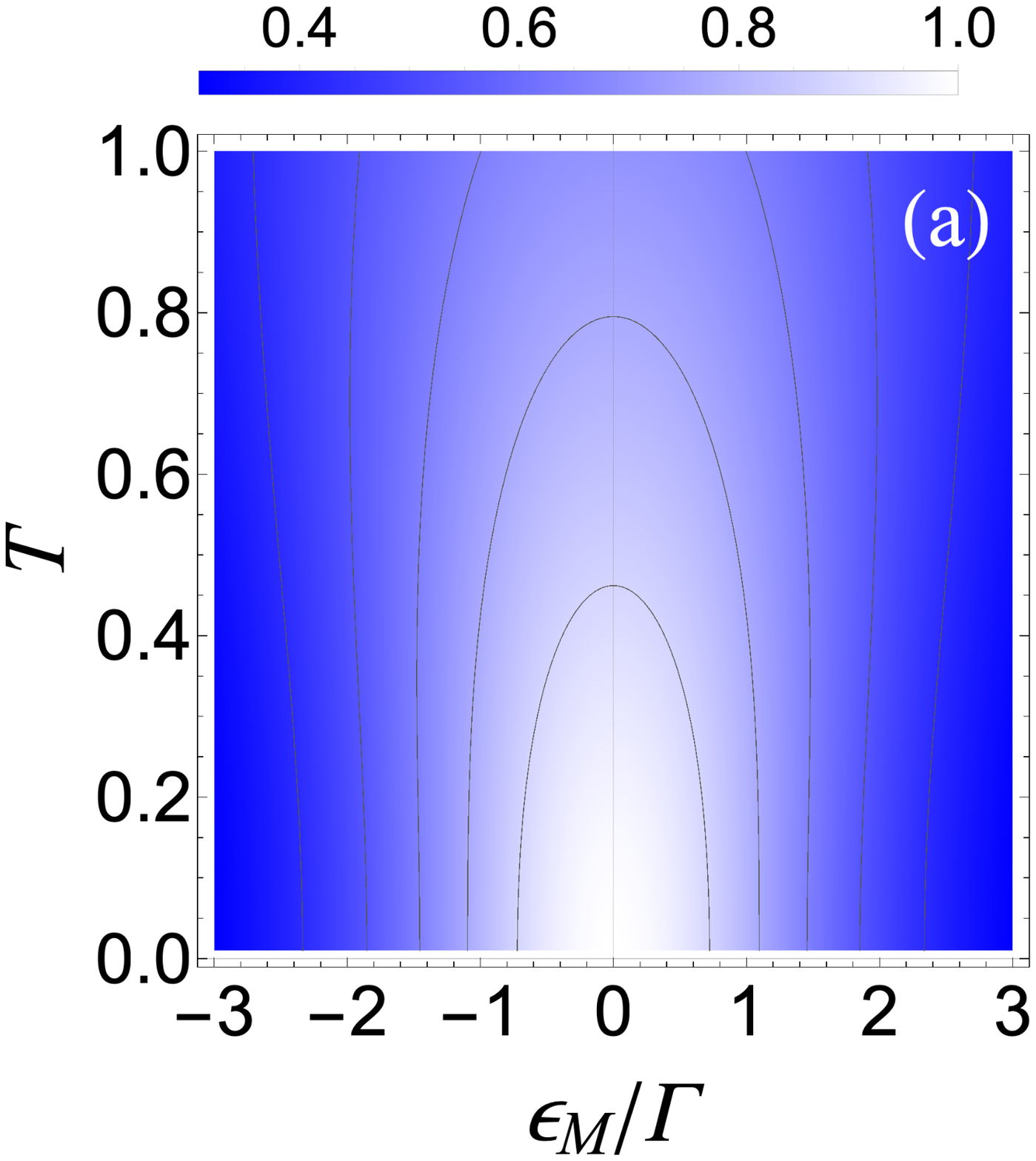}\includegraphics[scale=0.28]{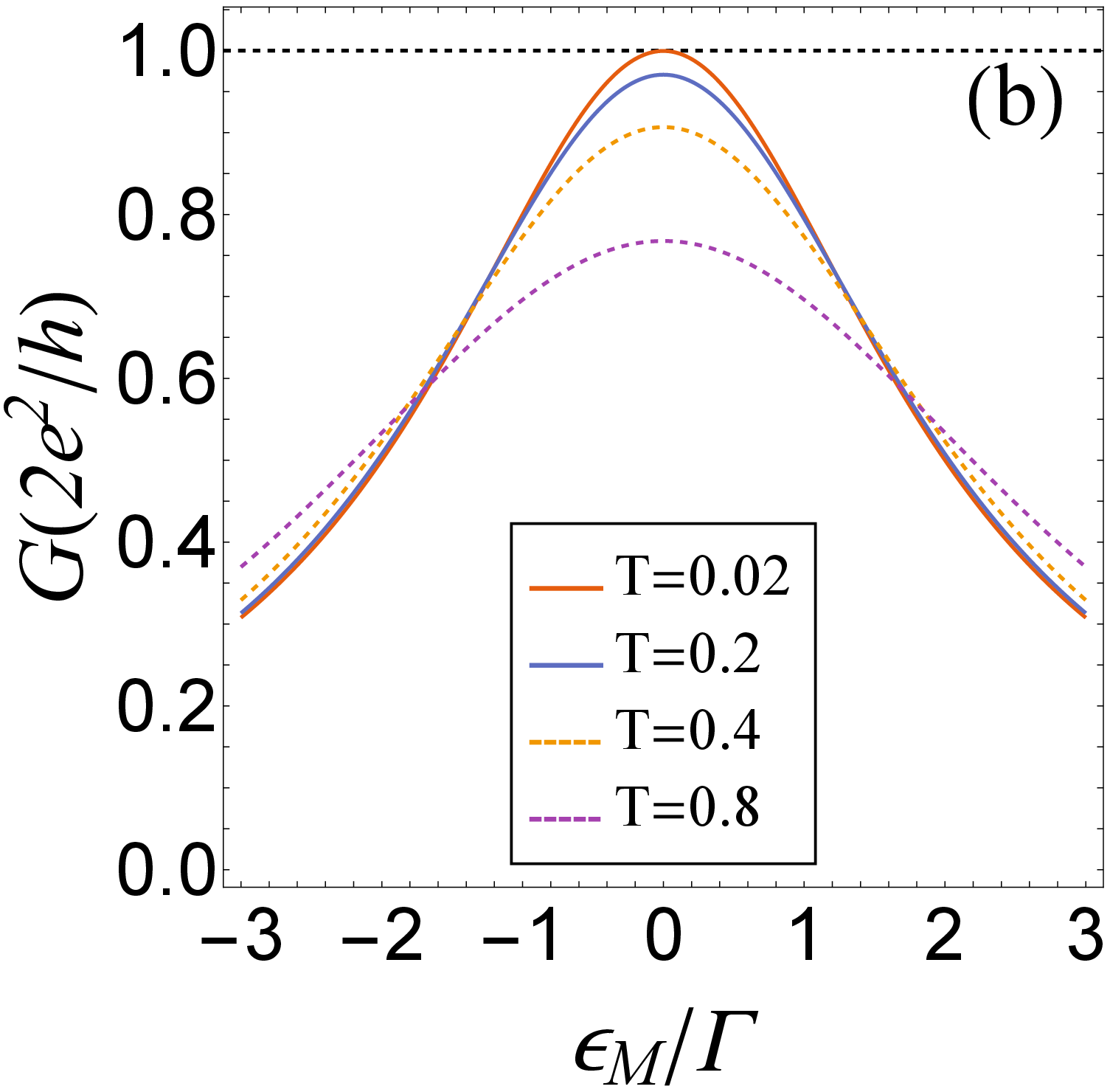}

\includegraphics[scale=0.28]{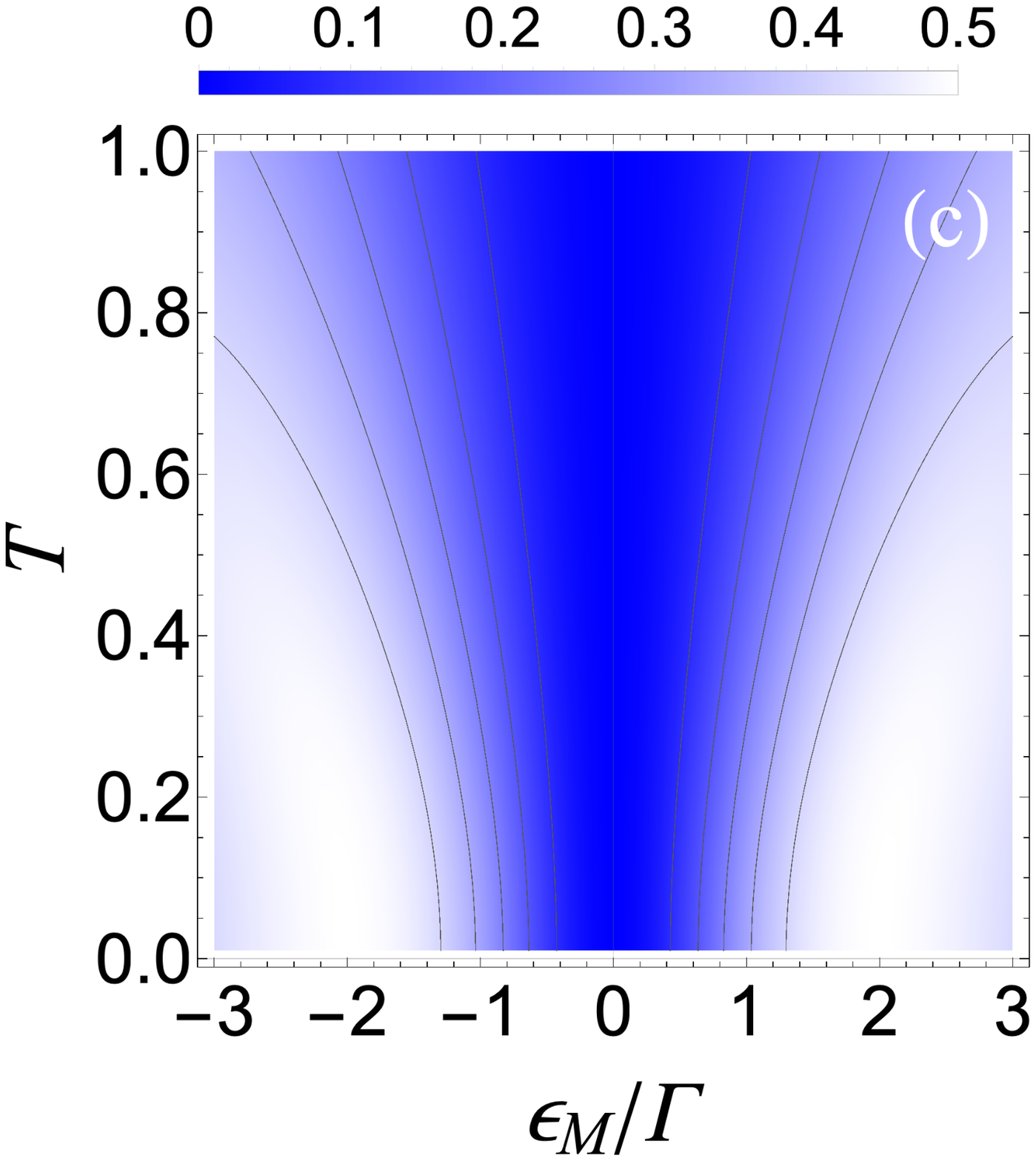}\includegraphics[scale=0.28]{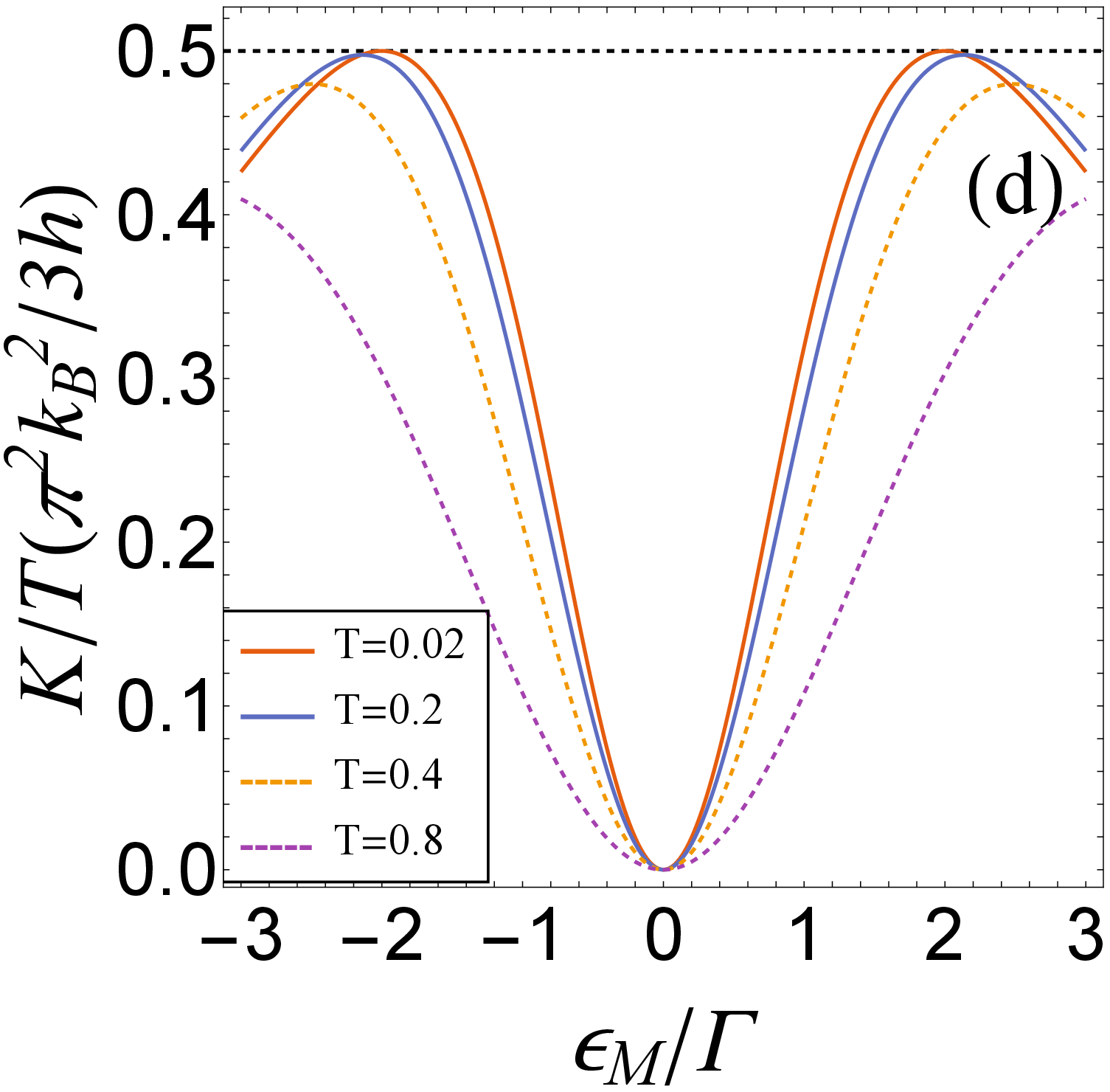}

\caption{The electrical conductance $G_{11}$ and thermal conductance $K/T$
for two MZMs. The coupling strength is set to $\Gamma=1.$  (a) Electrical
conductance $G_{11}$ as a function of the gap $\epsilon_{M}/\Gamma$
and the temperature $T.$ (b) Cross sections of the electrical conductance
at various temperatures $T=0.02$, $0.2$, $0.4,$ and $0.8.$ (c)
Thermal conductance $K/T$ as a function of the gap $\epsilon_{M}/\Gamma$
and the temperature $T.$ (d) Cross sections of the thermal conductance
at various temperatures $T=0.02$, $0.2$, $0.4,$ and $0.8.$ \label{fig:conduc_twoMZM_}}
\end{figure}

\begin{figure}
\includegraphics[scale=0.25]{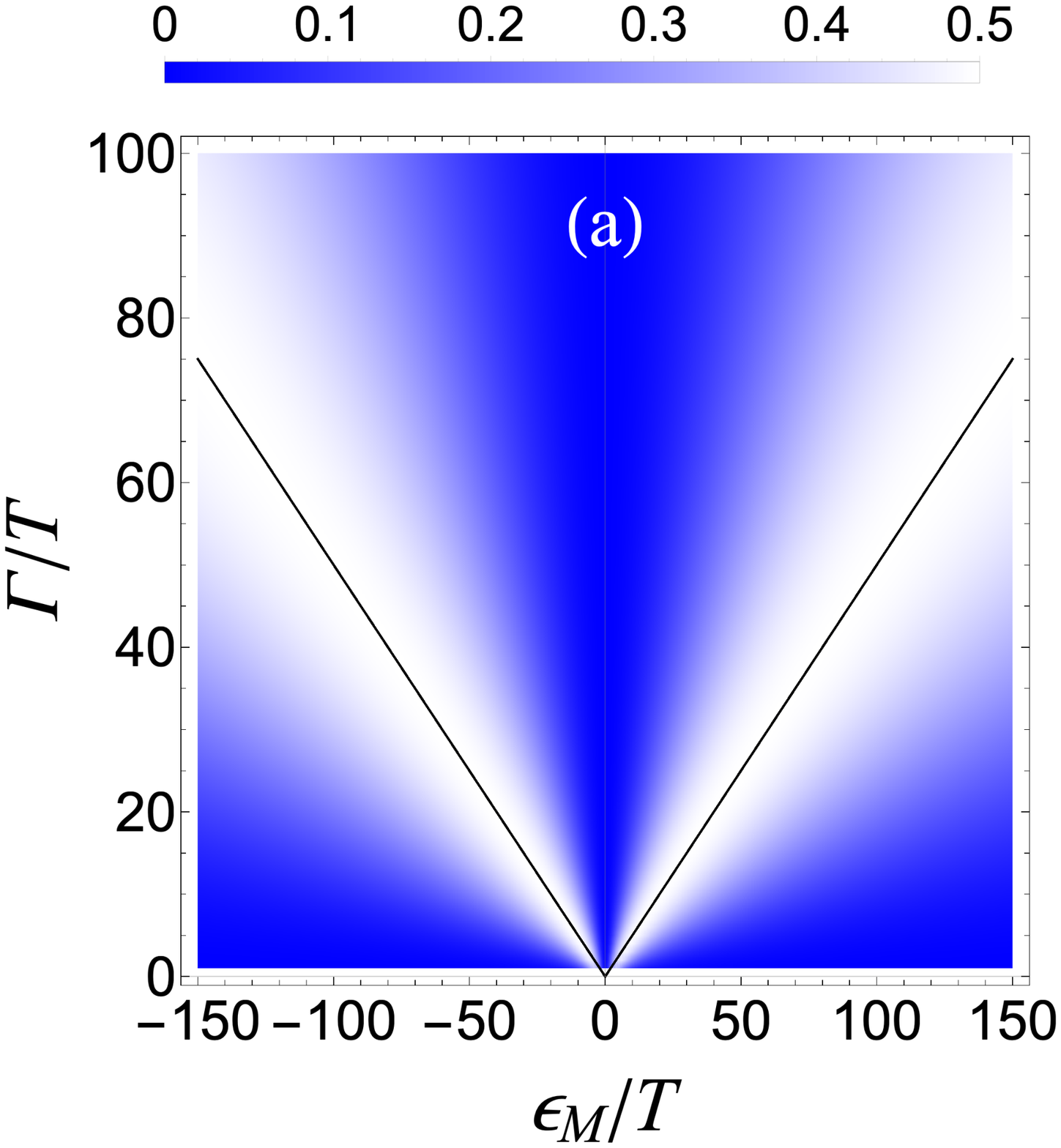}\includegraphics[scale=0.25]{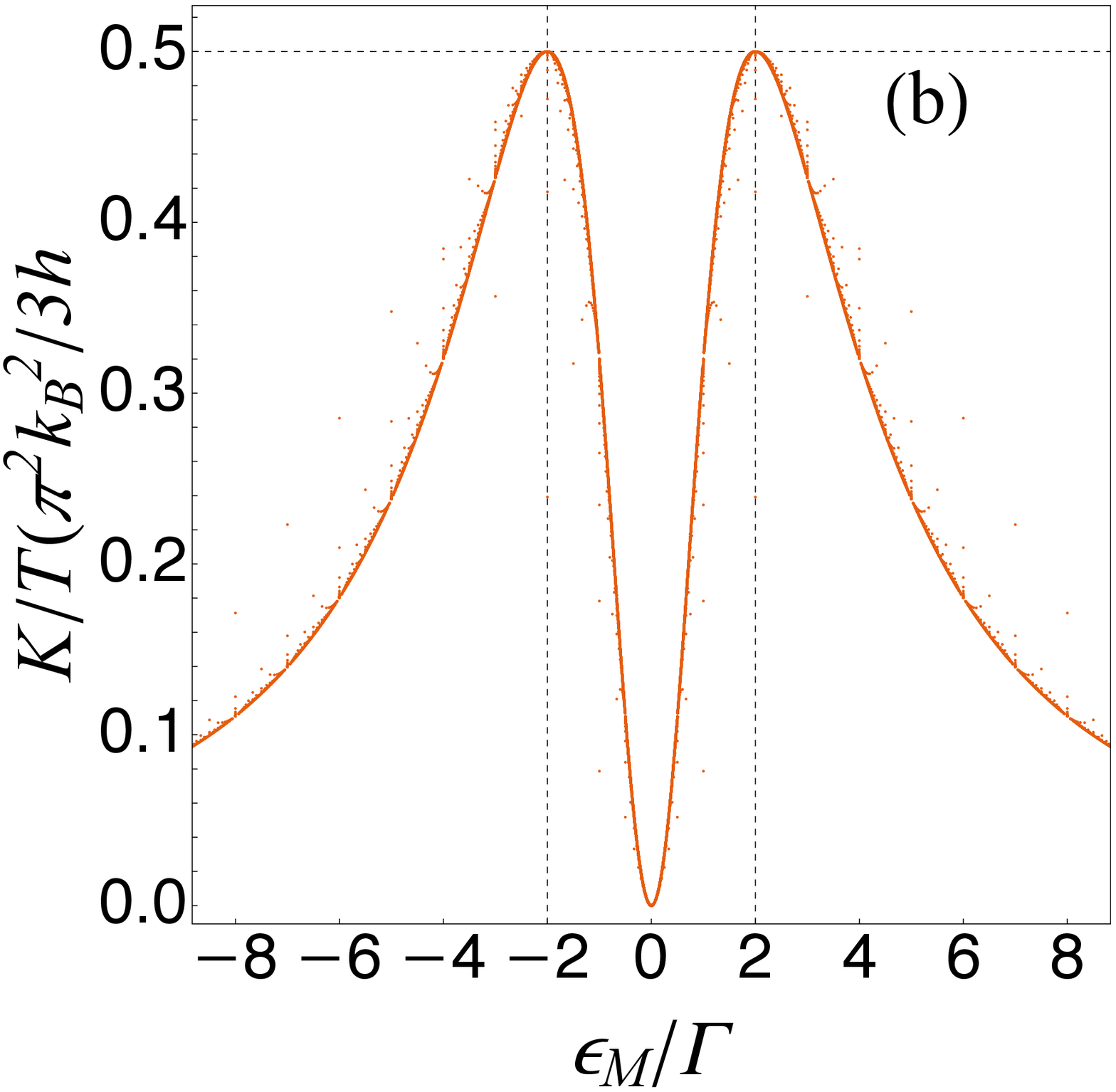}

\caption{(a) The thermal conductance $K/T$ as a function of $\Gamma/T$ and
$\epsilon_{M}/T$. The solid line is $2\Gamma=\epsilon_{M}.$ (b)
The thermal conductance $K/T$ as a function of $\epsilon_{M}/\Gamma$
at $T=0.02$. The maximum of $K/T$ is half of a thermal conductance
quantum $\pi^{2}k_{B}^{2}/6h,$ which is obtained in the limit $\epsilon_{M}\to\infty$,
$\Gamma\to\infty$ and $\epsilon_{M}/\Gamma=2.$ The points deviating
from the red line correspond to small $\epsilon_{M}$ and small $\Gamma$.
\label{fig:thermal_conduc_behavior}}
\end{figure}

In Fig.$\;$\ref{fig:conduc_twoMZM_}, we show $K(\epsilon_{M},T)$
and $G(\epsilon_{M},T).$ The behavior of $G$ is consistent with
previous studies: it is quantized at $2e^{2}/h$ at zero gap $\epsilon_{M}=0$
at low temperature. The thermal conductance differs substantially.
It vanishes at zero gap regardless of the temperature, and increases
to the maximum at finite gap. Interestingly, the maximum of $K/T$
is about half thermal conductance quantum $(1/2)\pi^{2}k_{B}^{2}/3h$.
As the temperature increases, the quantization is smeared out gradually.
In the following, we demonstrate that the quantization of $K/T$ is
in fact exact. We measure the energy in the unit of $T,$ i.e., we
scale $\omega\to\beta\omega$, $\Gamma\to\beta\Gamma,$ $\epsilon_{M}\to\beta\epsilon_{M}$.
Then $K$ can be written as 
\[
K=T\intw\;2\omega^{2}\bbT_{E}(\omega)\frac{1}{4\cosh^{2}\frac{\omega}{2}}.
\]
The integral reaches its maximum in the limit $\Gamma\to\infty$ and
$\epsilon_{M}\to\infty$. In this limit, we can approximate the integrand
$\bbT_{E}(\omega)\approx\bbT_{E}(0)$ and carry out the integral 
\begin{align*}
K/T & =\bbT_{E}(0)\intw\frac{\omega^{2}}{2\cosh^{2}\omega/2}=\frac{\pi}{3}\bbT_{E}(0)\\
 & =\frac{\pi}{3}\frac{4\Gamma^{2}\epsilon_{M}^{2}}{\left(4\Gamma^{2}+\epsilon_{M}^{2}\right)^{2}}\leq\frac{\pi}{12}=\half\frac{\pi^{2}k_{B}^{2}}{3h},
\end{align*}
 where the equality is obtained at $\epsilon_{M}/\Gamma=2$ and we
resort to SI unit in the last equality. We show $K/T$ as a function
of $\Gamma/T$ and $\epsilon_{M}/T$ in Fig.$\;$\ref{fig:thermal_conduc_behavior}.

According to Eq.$\;$(\ref{eq:work_power_def}), the work power is
\[
\dot{W}=J_{1}^{Q}+J_{2}^{Q}=-\mu_{1}J_{1}^{N}-\mu_{2}J_{2}^{N}=-G(\mu_{1}^{2}+\mu_{2}^{2}),
\]
which is always negative. The heat current is 
\begin{align*}
J_{1}^{Q} & =\frac{L_{3,3}}{T^{2}}\delta T,\quad J_{2}^{Q}=-\frac{L_{3,3}}{T^{2}}\delta T.
\end{align*}
Assume $\delta T>0$, then $J_{1}^{Q}>0$ and $J_{2}^{Q}<0$. It indicates
that the two MZMs as a thermoelectric device, always consumes energy
and cannot operate as a heat engine.

\subsection{Nonlinear transport}

The explicit expression of the currents has a consequence on the
response coefficients. The occupation number 
\begin{equation}
n_{1e}(\omega)-n_{1h}(\omega)=\frac{\sinh A_{1}}{\cosh\left[(\beta_{2}-A_{3})\omega\right]+\cosh A_{1}}\label{eq:n1e_minus_n1h}
\end{equation}
is an odd function of $A_{1},$ thus all responses coefficients corresponding
to even power term of $A_{1}$ vanishes. Eq.$\;$(\ref{eq:n1e_minus_n1h})
is also an even function of $\omega.$ Differentiating with respect
to $A_{3}$ won't change the parity of Eq.$\;$(\ref{eq:n1e_minus_n1h}).
So the responses coefficients only correspond to odd powers of $A_{1}$
are nonzero. Similar consideration applies to $J_{2}^{N}.$ Since
$J_{2}^{N}$ does not depend on $A_{3}$, the response coefficients
of $J_{2}^{N}$ are diagonal. From the expression of currents Eqs.$\;$(\ref{eq:current_1N_MZM},\ref{eq:current_2N_MZM}),
we find the second-order response coefficients
\begin{align*}
M_{1,13} & =M_{1,31}=\intw\;2\omega\bbT_{N}p_{0}(1-p_{0})(1-2p_{0}),\\
M_{3,33} & =\intw\;2\omega^{3}\bbT_{N}p_{0}(1-p_{0})(1-2p_{0}),\\
M_{3,22} & =\intw\;(-2\omega)\bbT_{N}p_{0}(1-p_{0})(1-2p_{0}),\\
M_{3,11} & =\intw\;2\omega\bbT_{E}p_{0}(1-p_{0})(1-2p_{0}),
\end{align*}
where $p_{0}=1/(e^{\beta_{2}\omega}+1)$. All other second-order coefficients
vanish.

The diffusivities are (due to the symmetry of the index, only six
of them are independent)
\begin{widetext}
\begin{align*}
D_{11}(\bmA) & =\half\intw[\bbT_{1}(n_{1e}\barn_{2e}+\barn_{1e}n_{2e}+n_{1e}\barn_{2h}+\barn_{1e}n_{2h})+2\bbC_{4}\bbT_{3}(n_{1e}\barn_{1h}+\barn_{1e}n_{1h})-(j_{1}^{N})^{2}],\\
D_{22}(\bmA) & =\half\intw[\bbT_{1}(n_{1e}\barn_{2e}+\barn_{1e}n_{2e}+n_{1e}\barn_{2h}+\barn_{1e}n_{2h})+2\bbC_{3}\bbT_{4}(n_{2e}\barn_{2h}+\barn_{2e}n_{2h})-(j_{2}^{N})^{2}],\\
D_{33}(\bmA) & =\half\intw[\omega^{2}\bbT_{1}(n_{1e}\barn_{2e}+\barn_{1e}n_{2e}+n_{1e}\barn_{2h}+\barn_{1e}n_{2h})-(j_{1}^{E})^{2}],\\
D_{12}(\bmA) & =\half\intw[\bbT_{1}(-n_{1e}\barn_{2e}-\barn_{1e}n_{2e}+n_{1e}\barn_{2h}+\barn_{1e}n_{2h})+2\bbT_{3}\bbT_{4}(n_{1e}-n_{1h})(n_{2e}-n_{2h})-j_{1}^{N}j_{2}^{N}],\\
D_{13}(\bmA) & =\half\intw\omega[\bbT_{1}(n_{1e}\barn_{2e}+\barn_{1e}n_{2e}+n_{1e}\barn_{2h}+\barn_{1e}n_{2h})-j_{1}^{N}j_{1}^{E}],\\
D_{23}(\bmA) & =\half\intw\omega[\bbT_{1}(-n_{1e}\barn_{2e}-\barn_{1e}n_{2e}+n_{1e}\barn_{2h}+\barn_{1e}n_{2h})-j_{2}^{N}j_{1}^{E}].
\end{align*}
At zero affinity, they reduce to 
\begin{align*}
D_{11}(\bmA=0) & =\intw\;2\bbT_{N}p_{0}(1-p_{0})=L_{1,1},\\
D_{22}(\bmA=0) & =\intw\;2\bbT_{N}p_{0}(1-p_{0})=L_{2,2},\\
D_{33}(\bmA=0) & =\intw\;2\omega\bbT_{N}p_{0}(1-p_{0})=L_{3,3},\\
D_{13}(\bmA=0) & =\intw\;2\omega\bbT_{E}p_{0}(1-p_{0})=L_{1,3},\\
D_{12}(\bmA=0) & =D_{23}(\bmA=0)=0
\end{align*}
as expected. We find that although the mean currents of the left and
the right reservoirs are decoupled, the diffusivity encodes the information
of the two LARs. The derivatives of diffusivity at zero affinities
are 
\begin{align*}
\frac{\partial D_{1i}}{\partial A_{j}} & =\intw\;p_{0}(1-p_{0})(1-2p_{0})\begin{pmatrix}0 & 0 & \omega(\bbT_{1}+2\bbC_{4}\bbT_{3})\\
0 & 0 & 0\\
\omega\bbT_{1} & 0 & 0
\end{pmatrix},\\
\frac{\partial D_{2i}}{\partial A_{j}} & =\intw\;p_{0}(1-p_{0})(1-2p_{0})\begin{pmatrix}0 & 0 & 0\\
0 & 0 & \omega\bbT_{1}\\
0 & -\omega\bbT_{1} & 0
\end{pmatrix},\\
\frac{\partial D_{3i}}{\partial A_{j}} & =\intw\;p_{0}(1-p_{0})(1-2p_{0})\begin{pmatrix}\omega\bbT_{1} & 0 & 0\\
0 & -\omega\bbT_{1} & 0\\
0 & 0 & \omega^{3}\bbT_{1}
\end{pmatrix}.
\end{align*}
The symmetric sum of $D_{ij,k}$ is 
\begin{align*}
\frac{\partial D_{1i}}{\partial A_{j}}+\frac{\partial D_{1j}}{\partial A_{i}} & =\begin{pmatrix}0 & 0 & D_{11,3}+D_{13,1}\\
0 & 0 & 0\\
D_{13,1}+D_{11,3} & 0 & 0
\end{pmatrix},\\
\frac{\partial D_{2i}}{\partial A_{j}}+\frac{\partial D_{2j}}{\partial A_{i}} & =0,\\
\frac{\partial D_{3i}}{\partial A_{j}}+\frac{\partial D_{3j}}{\partial A_{i}} & =\begin{pmatrix}2D_{13,1} & 0 & 0\\
0 & 2D_{23,2} & 0\\
0 & 0 & 2D_{33,3}
\end{pmatrix}.
\end{align*}
Compare to the expression of $M_{i,jk}$ we verify Eq.$\;$(\ref{eq:sec_order_response_relation})
\[
M_{i,jk}=\lrp{\frac{\partial D_{ij}}{\partial A_{k}}+\frac{\partial D_{ik}}{\partial A_{j}}}\bigg|_{\bfA=0}.
\]
\end{widetext}

\appendix
\bibliographystyle{apsrev4-1}
\bibliography{ref}

\providecommand{\noopsort}[1]{}\providecommand{\singleletter}[1]{#1}%
\begin{thebibliography}{46}%
\makeatletter
\providecommand \@ifxundefined [1]{%
 \@ifx{#1\undefined}
}%
\providecommand \@ifnum [1]{%
 \ifnum #1\expandafter \@firstoftwo
 \else \expandafter \@secondoftwo
 \fi
}%
\providecommand \@ifx [1]{%
 \ifx #1\expandafter \@firstoftwo
 \else \expandafter \@secondoftwo
 \fi
}%
\providecommand \natexlab [1]{#1}%
\providecommand \enquote  [1]{``#1''}%
\providecommand \bibnamefont  [1]{#1}%
\providecommand \bibfnamefont [1]{#1}%
\providecommand \citenamefont [1]{#1}%
\providecommand \href@noop [0]{\@secondoftwo}%
\providecommand \href [0]{\begingroup \@sanitize@url \@href}%
\providecommand \@href[1]{\@@startlink{#1}\@@href}%
\providecommand \@@href[1]{\endgroup#1\@@endlink}%
\providecommand \@sanitize@url [0]{\catcode `\\12\catcode `\$12\catcode
  `\&12\catcode `\#12\catcode `\^12\catcode `\_12\catcode `\%12\relax}%
\providecommand \@@startlink[1]{}%
\providecommand \@@endlink[0]{}%
\providecommand \url  [0]{\begingroup\@sanitize@url \@url }%
\providecommand \@url [1]{\endgroup\@href {#1}{\urlprefix }}%
\providecommand \urlprefix  [0]{URL }%
\providecommand \Eprint [0]{\href }%
\providecommand \doibase [0]{http://dx.doi.org/}%
\providecommand \selectlanguage [0]{\@gobble}%
\providecommand \bibinfo  [0]{\@secondoftwo}%
\providecommand \bibfield  [0]{\@secondoftwo}%
\providecommand \translation [1]{[#1]}%
\providecommand \BibitemOpen [0]{}%
\providecommand \bibitemStop [0]{}%
\providecommand \bibitemNoStop [0]{.\EOS\space}%
\providecommand \EOS [0]{\spacefactor3000\relax}%
\providecommand \BibitemShut  [1]{\csname bibitem#1\endcsname}%
\let\auto@bib@innerbib\@empty
%</preamble>
\bibitem [{\citenamefont {Onsager}(1931{\natexlab{a}})}]{Onsager1931PR}%
  \BibitemOpen
  \bibfield  {author} {\bibinfo {author} {\bibfnamefont {L.}~\bibnamefont
  {Onsager}},\ }\href {\doibase 10.1103/PhysRev.37.405} {\bibfield  {journal}
  {\bibinfo  {journal} {Phys. Rev.}\ }\textbf {\bibinfo {volume} {37}},\
  \bibinfo {pages} {405} (\bibinfo {year} {1931}{\natexlab{a}})}\BibitemShut
  {NoStop}%
\bibitem [{\citenamefont {Onsager}(1931{\natexlab{b}})}]{Onsager1931PRa}%
  \BibitemOpen
  \bibfield  {author} {\bibinfo {author} {\bibfnamefont {L.}~\bibnamefont
  {Onsager}},\ }\href {\doibase 10.1103/PhysRev.38.2265} {\bibfield  {journal}
  {\bibinfo  {journal} {Phys. Rev.}\ }\textbf {\bibinfo {volume} {38}},\
  \bibinfo {pages} {2265} (\bibinfo {year} {1931}{\natexlab{b}})}\BibitemShut
  {NoStop}%
\bibitem [{\citenamefont {Casimir}(1945)}]{Casimir1945RMP}%
  \BibitemOpen
  \bibfield  {author} {\bibinfo {author} {\bibfnamefont {H.~B.~G.}\
  \bibnamefont {Casimir}},\ }\href {\doibase 10.1103/RevModPhys.17.343}
  {\bibfield  {journal} {\bibinfo  {journal} {Rev. Mod. Phys.}\ }\textbf
  {\bibinfo {volume} {17}},\ \bibinfo {pages} {343} (\bibinfo {year}
  {1945})}\BibitemShut {NoStop}%
\bibitem [{\citenamefont {Callen}\ and\ \citenamefont
  {Welton}(1951)}]{Callen1951PR}%
  \BibitemOpen
  \bibfield  {author} {\bibinfo {author} {\bibfnamefont {H.~B.}\ \bibnamefont
  {Callen}}\ and\ \bibinfo {author} {\bibfnamefont {T.~A.}\ \bibnamefont
  {Welton}},\ }\href {\doibase 10.1103/PhysRev.83.34} {\bibfield  {journal}
  {\bibinfo  {journal} {Phys. Rev.}\ }\textbf {\bibinfo {volume} {83}},\
  \bibinfo {pages} {34} (\bibinfo {year} {1951})}\BibitemShut {NoStop}%
\bibitem [{\citenamefont {Evans}\ \emph {et~al.}(1993)\citenamefont {Evans},
  \citenamefont {Cohen},\ and\ \citenamefont {Morriss}}]{Evans1993prl}%
  \BibitemOpen
  \bibfield  {author} {\bibinfo {author} {\bibfnamefont {D.~J.}\ \bibnamefont
  {Evans}}, \bibinfo {author} {\bibfnamefont {E.~G.~D.}\ \bibnamefont {Cohen}},
  \ and\ \bibinfo {author} {\bibfnamefont {G.~P.}\ \bibnamefont {Morriss}},\
  }\href {\doibase 10.1103/PhysRevLett.71.2401} {\bibfield  {journal} {\bibinfo
   {journal} {Phys. Rev. Lett.}\ }\textbf {\bibinfo {volume} {71}},\ \bibinfo
  {pages} {2401} (\bibinfo {year} {1993})}\BibitemShut {NoStop}%
\bibitem [{\citenamefont {Evans}\ and\ \citenamefont
  {Searles}(1994)}]{evans1994pre}%
  \BibitemOpen
  \bibfield  {author} {\bibinfo {author} {\bibfnamefont {D.~J.}\ \bibnamefont
  {Evans}}\ and\ \bibinfo {author} {\bibfnamefont {D.~J.}\ \bibnamefont
  {Searles}},\ }\href@noop {} {\bibfield  {journal} {\bibinfo  {journal} {Phys.
  Rev. E}\ }\textbf {\bibinfo {volume} {50}},\ \bibinfo {pages} {1645}
  (\bibinfo {year} {1994})}\BibitemShut {NoStop}%
\bibitem [{\citenamefont {Gallavotti}\ and\ \citenamefont
  {Cohen}(1995{\natexlab{a}})}]{Gallavotti1995jsp}%
  \BibitemOpen
  \bibfield  {author} {\bibinfo {author} {\bibfnamefont {G.}~\bibnamefont
  {Gallavotti}}\ and\ \bibinfo {author} {\bibfnamefont {E.~G.~D.}\ \bibnamefont
  {Cohen}},\ }\href@noop {} {\bibfield  {journal} {\bibinfo  {journal} {J.
  Stat. Phys.}\ }\textbf {\bibinfo {volume} {80}},\ \bibinfo {pages} {931}
  (\bibinfo {year} {1995}{\natexlab{a}})}\BibitemShut {NoStop}%
\bibitem [{\citenamefont {Gallavotti}\ and\ \citenamefont
  {Cohen}(1995{\natexlab{b}})}]{Gallavotti1995prl}%
  \BibitemOpen
  \bibfield  {author} {\bibinfo {author} {\bibfnamefont {G.}~\bibnamefont
  {Gallavotti}}\ and\ \bibinfo {author} {\bibfnamefont {E.~G.~D.}\ \bibnamefont
  {Cohen}},\ }\href {\doibase 10.1103/PhysRevLett.74.2694} {\bibfield
  {journal} {\bibinfo  {journal} {Phys. Rev. Lett.}\ }\textbf {\bibinfo
  {volume} {74}},\ \bibinfo {pages} {2694} (\bibinfo {year}
  {1995}{\natexlab{b}})}\BibitemShut {NoStop}%
\bibitem [{\citenamefont {Kurchan}(1998)}]{Kurchan1998jpa}%
  \BibitemOpen
  \bibfield  {author} {\bibinfo {author} {\bibfnamefont {J.}~\bibnamefont
  {Kurchan}},\ }\href {\doibase 10.1088/0305-4470/31/16/003} {\bibfield
  {journal} {\bibinfo  {journal} {J. Phys. A: Math. Gen.}\ }\textbf {\bibinfo
  {volume} {31}},\ \bibinfo {pages} {3719} (\bibinfo {year}
  {1998})}\BibitemShut {NoStop}%
\bibitem [{\citenamefont {Maes}(1999)}]{Maes1999jsp}%
  \BibitemOpen
  \bibfield  {author} {\bibinfo {author} {\bibfnamefont {C.}~\bibnamefont
  {Maes}},\ }\href {\doibase 10.1023/A:1004541830999} {\bibfield  {journal}
  {\bibinfo  {journal} {J. Stat. Phys.}\ }\textbf {\bibinfo {volume} {95}},\
  \bibinfo {pages} {367} (\bibinfo {year} {1999})}\BibitemShut {NoStop}%
\bibitem [{\citenamefont {Lebowitz}\ and\ \citenamefont
  {Spohn}(1999)}]{Lebowitz1999jsp}%
  \BibitemOpen
  \bibfield  {author} {\bibinfo {author} {\bibfnamefont {J.~L.}\ \bibnamefont
  {Lebowitz}}\ and\ \bibinfo {author} {\bibfnamefont {H.}~\bibnamefont
  {Spohn}},\ }\href {\doibase 10.1023/A:1004589714161} {\bibfield  {journal}
  {\bibinfo  {journal} {J. Stat. Phys.}\ }\textbf {\bibinfo {volume} {95}},\
  \bibinfo {pages} {333} (\bibinfo {year} {1999})}\BibitemShut {NoStop}%
\bibitem [{\citenamefont {Esposito}\ \emph {et~al.}(2009)\citenamefont
  {Esposito}, \citenamefont {Harbola},\ and\ \citenamefont
  {Mukamel}}]{Esposito2009rmp}%
  \BibitemOpen
  \bibfield  {author} {\bibinfo {author} {\bibfnamefont {M.}~\bibnamefont
  {Esposito}}, \bibinfo {author} {\bibfnamefont {U.}~\bibnamefont {Harbola}}, \
  and\ \bibinfo {author} {\bibfnamefont {S.}~\bibnamefont {Mukamel}},\ }\href
  {\doibase 10.1103/RevModPhys.81.1665} {\bibfield  {journal} {\bibinfo
  {journal} {Rev. Mod. Phys.}\ }\textbf {\bibinfo {volume} {81}},\ \bibinfo
  {pages} {1665} (\bibinfo {year} {2009})}\BibitemShut {NoStop}%
\bibitem [{\citenamefont {Campisi}\ \emph {et~al.}(2011)\citenamefont
  {Campisi}, \citenamefont {H\"anggi},\ and\ \citenamefont
  {Talkner}}]{Campisi2011rmp}%
  \BibitemOpen
  \bibfield  {author} {\bibinfo {author} {\bibfnamefont {M.}~\bibnamefont
  {Campisi}}, \bibinfo {author} {\bibfnamefont {P.}~\bibnamefont {H\"anggi}}, \
  and\ \bibinfo {author} {\bibfnamefont {P.}~\bibnamefont {Talkner}},\ }\href
  {\doibase 10.1103/RevModPhys.83.771} {\bibfield  {journal} {\bibinfo
  {journal} {Rev. Mod. Phys.}\ }\textbf {\bibinfo {volume} {83}},\ \bibinfo
  {pages} {771} (\bibinfo {year} {2011})}\BibitemShut {NoStop}%
\bibitem [{\citenamefont {Jarzynski}(2011)}]{Jarzynski2011an}%
  \BibitemOpen
  \bibfield  {author} {\bibinfo {author} {\bibfnamefont {C.}~\bibnamefont
  {Jarzynski}},\ }\href {\doibase 10.1146/annurev-conmatphys-062910-140506}
  {\bibfield  {journal} {\bibinfo  {journal} {Annu. Rev. Condens. Matter
  Phys.}\ }\textbf {\bibinfo {volume} {2}},\ \bibinfo {pages} {329} (\bibinfo
  {year} {2011})}\BibitemShut {NoStop}%
\bibitem [{\citenamefont {Seifert}(2012)}]{Seifert2012rpp}%
  \BibitemOpen
  \bibfield  {author} {\bibinfo {author} {\bibfnamefont {U.}~\bibnamefont
  {Seifert}},\ }\href {\doibase 10.1088/0034-4885/75/12/126001} {\bibfield
  {journal} {\bibinfo  {journal} {Rep. Prog. Phys.}\ }\textbf {\bibinfo
  {volume} {75}},\ \bibinfo {pages} {126001} (\bibinfo {year}
  {2012})}\BibitemShut {NoStop}%
\bibitem [{\citenamefont {Saito}\ and\ \citenamefont
  {Utsumi}(2008)}]{Saito2008prb}%
  \BibitemOpen
  \bibfield  {author} {\bibinfo {author} {\bibfnamefont {K.}~\bibnamefont
  {Saito}}\ and\ \bibinfo {author} {\bibfnamefont {Y.}~\bibnamefont {Utsumi}},\
  }\href {\doibase 10.1103/PhysRevB.78.115429} {\bibfield  {journal} {\bibinfo
  {journal} {Phys. Rev. B}\ }\textbf {\bibinfo {volume} {78}},\ \bibinfo
  {pages} {115429} (\bibinfo {year} {2008})}\BibitemShut {NoStop}%
\bibitem [{\citenamefont {Andrieux}\ and\ \citenamefont
  {Gaspard}(2008)}]{Andrieux2008prl}%
  \BibitemOpen
  \bibfield  {author} {\bibinfo {author} {\bibfnamefont {D.}~\bibnamefont
  {Andrieux}}\ and\ \bibinfo {author} {\bibfnamefont {P.}~\bibnamefont
  {Gaspard}},\ }\href {\doibase 10.1103/PhysRevLett.100.230404} {\bibfield
  {journal} {\bibinfo  {journal} {Phys. Rev. Lett.}\ }\textbf {\bibinfo
  {volume} {100}},\ \bibinfo {pages} {230404} (\bibinfo {year}
  {2008})}\BibitemShut {NoStop}%
\bibitem [{\citenamefont {Andrieux}\ \emph {et~al.}(2009)\citenamefont
  {Andrieux}, \citenamefont {Gaspard}, \citenamefont {Monnai},\ and\
  \citenamefont {Tasaki}}]{Andrieux2009njp}%
  \BibitemOpen
  \bibfield  {author} {\bibinfo {author} {\bibfnamefont {D.}~\bibnamefont
  {Andrieux}}, \bibinfo {author} {\bibfnamefont {P.}~\bibnamefont {Gaspard}},
  \bibinfo {author} {\bibfnamefont {T.}~\bibnamefont {Monnai}}, \ and\ \bibinfo
  {author} {\bibfnamefont {S.}~\bibnamefont {Tasaki}},\ }\href {\doibase
  10.1088/1367-2630/11/4/043014} {\bibfield  {journal} {\bibinfo  {journal}
  {New J. Phys.}\ }\textbf {\bibinfo {volume} {11}},\ \bibinfo {pages} {043014}
  (\bibinfo {year} {2009})}\BibitemShut {NoStop}%
\bibitem [{\citenamefont {Gaspard}(2013)}]{Gaspard2013njp}%
  \BibitemOpen
  \bibfield  {author} {\bibinfo {author} {\bibfnamefont {P.}~\bibnamefont
  {Gaspard}},\ }\href {\doibase 10.1088/1367-2630/15/11/115014} {\bibfield
  {journal} {\bibinfo  {journal} {New J. Phys.}\ }\textbf {\bibinfo {volume}
  {15}},\ \bibinfo {pages} {115014} (\bibinfo {year} {2013})}\BibitemShut
  {NoStop}%
\bibitem [{\citenamefont {Barbier}\ and\ \citenamefont
  {Gaspard}(2018)}]{Barbier2018jpa}%
  \BibitemOpen
  \bibfield  {author} {\bibinfo {author} {\bibfnamefont {M.}~\bibnamefont
  {Barbier}}\ and\ \bibinfo {author} {\bibfnamefont {P.}~\bibnamefont
  {Gaspard}},\ }\href {\doibase 10.1088/1751-8121/aaf218} {\bibfield  {journal}
  {\bibinfo  {journal} {Journal of Physics A: Mathematical and Theoretical}\
  }\textbf {\bibinfo {volume} {52}},\ \bibinfo {pages} {025003} (\bibinfo
  {year} {2018})}\BibitemShut {NoStop}%
\bibitem [{\citenamefont {Gu}\ and\ \citenamefont {Gaspard}(2019)}]{Gu2019pre}%
  \BibitemOpen
  \bibfield  {author} {\bibinfo {author} {\bibfnamefont {J.}~\bibnamefont
  {Gu}}\ and\ \bibinfo {author} {\bibfnamefont {P.}~\bibnamefont {Gaspard}},\
  }\href {\doibase 10.1103/PhysRevE.99.012137} {\bibfield  {journal} {\bibinfo
  {journal} {Phys. Rev. E}\ }\textbf {\bibinfo {volume} {99}},\ \bibinfo
  {pages} {012137} (\bibinfo {year} {2019})}\BibitemShut {NoStop}%
\bibitem [{\citenamefont {Barbier}\ and\ \citenamefont
  {Gaspard}(2020{\natexlab{a}})}]{Barbier2020JPAMT}%
  \BibitemOpen
  \bibfield  {author} {\bibinfo {author} {\bibfnamefont {M.}~\bibnamefont
  {Barbier}}\ and\ \bibinfo {author} {\bibfnamefont {P.}~\bibnamefont
  {Gaspard}},\ }\href {\doibase 10.1088/1751-8121/ab777e} {\bibfield  {journal}
  {\bibinfo  {journal} {J. Phys. A: Math. Theor.}\ }\textbf {\bibinfo {volume}
  {53}},\ \bibinfo {pages} {145002} (\bibinfo {year}
  {2020}{\natexlab{a}})}\BibitemShut {NoStop}%
\bibitem [{\citenamefont {Barbier}\ and\ \citenamefont
  {Gaspard}(2020{\natexlab{b}})}]{Barbier2020PRE}%
  \BibitemOpen
  \bibfield  {author} {\bibinfo {author} {\bibfnamefont {M.}~\bibnamefont
  {Barbier}}\ and\ \bibinfo {author} {\bibfnamefont {P.}~\bibnamefont
  {Gaspard}},\ }\href {\doibase 10.1103/PhysRevE.102.022141} {\bibfield
  {journal} {\bibinfo  {journal} {Phys. Rev. E}\ }\textbf {\bibinfo {volume}
  {102}},\ \bibinfo {pages} {022141} (\bibinfo {year}
  {2020}{\natexlab{b}})}\BibitemShut {NoStop}%
\bibitem [{\citenamefont {Gu}\ and\ \citenamefont {Gaspard}(2020)}]{Gu2020jsm}%
  \BibitemOpen
  \bibfield  {author} {\bibinfo {author} {\bibfnamefont {J.}~\bibnamefont
  {Gu}}\ and\ \bibinfo {author} {\bibfnamefont {P.}~\bibnamefont {Gaspard}},\
  }\href {\doibase 10.1088/1742-5468/abbcd5} {\bibfield  {journal} {\bibinfo
  {journal} {Journal of Statistical Mechanics: Theory and Experiment}\ }\textbf
  {\bibinfo {volume} {2020}},\ \bibinfo {pages} {103206} (\bibinfo {year}
  {2020})}\BibitemShut {NoStop}%
\bibitem [{\citenamefont {Wu}\ \emph {et~al.}(2022)\citenamefont {Wu},
  \citenamefont {Gu},\ and\ \citenamefont {Quan}}]{Hu2022arxiv}%
  \BibitemOpen
  \bibfield  {author} {\bibinfo {author} {\bibfnamefont {Y.-X.}\ \bibnamefont
  {Wu}}, \bibinfo {author} {\bibfnamefont {J.}~\bibnamefont {Gu}}, \ and\
  \bibinfo {author} {\bibfnamefont {H.~T.}\ \bibnamefont {Quan}},\ }\href
  {\doibase 10.48550/ARXIV.2205.08870} {\  (\bibinfo {year} {2022}),\
  10.48550/ARXIV.2205.08870}\BibitemShut {NoStop}%
\bibitem [{\citenamefont {Jarzynski}\ and\ \citenamefont
  {W\'ojcik}(2004)}]{Jarzynski2004prl}%
  \BibitemOpen
  \bibfield  {author} {\bibinfo {author} {\bibfnamefont {C.}~\bibnamefont
  {Jarzynski}}\ and\ \bibinfo {author} {\bibfnamefont {D.~K.}\ \bibnamefont
  {W\'ojcik}},\ }\href {\doibase 10.1103/PhysRevLett.92.230602} {\bibfield
  {journal} {\bibinfo  {journal} {Phys. Rev. Lett.}\ }\textbf {\bibinfo
  {volume} {92}},\ \bibinfo {pages} {230602} (\bibinfo {year}
  {2004})}\BibitemShut {NoStop}%
\bibitem [{\citenamefont {Zhang}\ and\ \citenamefont
  {Quan}(2021)}]{Zhang2021pre}%
  \BibitemOpen
  \bibfield  {author} {\bibinfo {author} {\bibfnamefont {F.}~\bibnamefont
  {Zhang}}\ and\ \bibinfo {author} {\bibfnamefont {H.~T.}\ \bibnamefont
  {Quan}},\ }\href {\doibase 10.1103/PhysRevE.103.032143} {\bibfield  {journal}
  {\bibinfo  {journal} {Phys. Rev. E}\ }\textbf {\bibinfo {volume} {103}},\
  \bibinfo {pages} {032143} (\bibinfo {year} {2021})}\BibitemShut {NoStop}%
\bibitem [{\citenamefont {Oreg}\ \emph {et~al.}(2010)\citenamefont {Oreg},
  \citenamefont {Refael},\ and\ \citenamefont {von Oppen}}]{Oreg2010prl}%
  \BibitemOpen
  \bibfield  {author} {\bibinfo {author} {\bibfnamefont {Y.}~\bibnamefont
  {Oreg}}, \bibinfo {author} {\bibfnamefont {G.}~\bibnamefont {Refael}}, \ and\
  \bibinfo {author} {\bibfnamefont {F.}~\bibnamefont {von Oppen}},\ }\href
  {\doibase 10.1103/PhysRevLett.105.177002} {\bibfield  {journal} {\bibinfo
  {journal} {Phys. Rev. Lett.}\ }\textbf {\bibinfo {volume} {105}},\ \bibinfo
  {pages} {177002} (\bibinfo {year} {2010})}\BibitemShut {NoStop}%
\bibitem [{\citenamefont {Lutchyn}\ \emph {et~al.}(2010)\citenamefont
  {Lutchyn}, \citenamefont {Sau},\ and\ \citenamefont
  {Das~Sarma}}]{Lutchyn2010prl}%
  \BibitemOpen
  \bibfield  {author} {\bibinfo {author} {\bibfnamefont {R.~M.}\ \bibnamefont
  {Lutchyn}}, \bibinfo {author} {\bibfnamefont {J.~D.}\ \bibnamefont {Sau}}, \
  and\ \bibinfo {author} {\bibfnamefont {S.}~\bibnamefont {Das~Sarma}},\ }\href
  {\doibase 10.1103/PhysRevLett.105.077001} {\bibfield  {journal} {\bibinfo
  {journal} {Phys. Rev. Lett.}\ }\textbf {\bibinfo {volume} {105}},\ \bibinfo
  {pages} {077001} (\bibinfo {year} {2010})}\BibitemShut {NoStop}%
\bibitem [{\citenamefont {Qiao}\ \emph {et~al.}(2021)\citenamefont {Qiao},
  \citenamefont {Li},\ and\ \citenamefont {Sun}}]{qiao2021arxiv}%
  \BibitemOpen
  \bibfield  {author} {\bibinfo {author} {\bibfnamefont {G.-J.}\ \bibnamefont
  {Qiao}}, \bibinfo {author} {\bibfnamefont {S.-W.}\ \bibnamefont {Li}}, \ and\
  \bibinfo {author} {\bibfnamefont {C.~P.}\ \bibnamefont {Sun}},\ }\href@noop
  {} {\bibfield  {journal} {\bibinfo  {journal} {ArXiv211213568 Cond-Mat}\ }
  (\bibinfo {year} {2021})},\ \Eprint {http://arxiv.org/abs/2112.13568}
  {arXiv:2112.13568 [cond-mat.mes-hall]} \BibitemShut {NoStop}%
\bibitem [{\citenamefont {Kamenev}(2011)}]{Kamenev2011book}%
  \BibitemOpen
  \bibfield  {author} {\bibinfo {author} {\bibfnamefont {A.}~\bibnamefont
  {Kamenev}},\ }\href {\doibase 10.1017/CBO9781139003667} {\emph {\bibinfo
  {title} {Field Theory of Non-Equilibrium Systems}}}\ (\bibinfo  {publisher}
  {Cambridge University Press},\ \bibinfo {year} {2011})\BibitemShut {NoStop}%
\bibitem [{\citenamefont {Shankar}(2017)}]{Shankar2017QFT}%
  \BibitemOpen
  \bibfield  {author} {\bibinfo {author} {\bibfnamefont {R.}~\bibnamefont
  {Shankar}},\ }\href@noop {} {\emph {\bibinfo {title} {Quantum Field Theory
  and Condensed Matter: An Introduction}}}\ (\bibinfo  {publisher} {Cambridge
  University Press},\ \bibinfo {year} {2017})\BibitemShut {NoStop}%
\bibitem [{\citenamefont {Nilsson}\ \emph {et~al.}(2008)\citenamefont
  {Nilsson}, \citenamefont {Akhmerov},\ and\ \citenamefont
  {Beenakker}}]{Nilsson2008prl}%
  \BibitemOpen
  \bibfield  {author} {\bibinfo {author} {\bibfnamefont {J.}~\bibnamefont
  {Nilsson}}, \bibinfo {author} {\bibfnamefont {A.~R.}\ \bibnamefont
  {Akhmerov}}, \ and\ \bibinfo {author} {\bibfnamefont {C.~W.~J.}\ \bibnamefont
  {Beenakker}},\ }\href {\doibase 10.1103/PhysRevLett.101.120403} {\bibfield
  {journal} {\bibinfo  {journal} {Phys. Rev. Lett.}\ }\textbf {\bibinfo
  {volume} {101}},\ \bibinfo {pages} {120403} (\bibinfo {year}
  {2008})}\BibitemShut {NoStop}%
\bibitem [{\citenamefont {Law}\ \emph {et~al.}(2009)\citenamefont {Law},
  \citenamefont {Lee},\ and\ \citenamefont {Ng}}]{Law2009prl}%
  \BibitemOpen
  \bibfield  {author} {\bibinfo {author} {\bibfnamefont {K.~T.}\ \bibnamefont
  {Law}}, \bibinfo {author} {\bibfnamefont {P.~A.}\ \bibnamefont {Lee}}, \ and\
  \bibinfo {author} {\bibfnamefont {T.~K.}\ \bibnamefont {Ng}},\ }\href
  {\doibase 10.1103/PhysRevLett.103.237001} {\bibfield  {journal} {\bibinfo
  {journal} {Phys. Rev. Lett.}\ }\textbf {\bibinfo {volume} {103}},\ \bibinfo
  {pages} {237001} (\bibinfo {year} {2009})}\BibitemShut {NoStop}%
\bibitem [{Note1()}]{Note1}%
  \BibitemOpen
  \bibinfo {note} {In fact, the CAR occurs when the site number is even. If the
  site number is odd, NT rather than CAR will occur.}\BibitemShut {Stop}%
\bibitem [{Note2()}]{Note2}%
  \BibitemOpen
  \bibinfo {note} {In fact, the number of sites should be larger than
  3.}\BibitemShut {Stop}%
\bibitem [{\citenamefont {Kitaev}(2001)}]{kitaev2001ps}%
  \BibitemOpen
  \bibfield  {author} {\bibinfo {author} {\bibfnamefont {A.~Y.}\ \bibnamefont
  {Kitaev}},\ }\href {\doibase 10.1070/1063-7869/44/10s/s29} {\bibfield
  {journal} {\bibinfo  {journal} {Physics-Uspekhi}\ }\textbf {\bibinfo {volume}
  {44}},\ \bibinfo {pages} {131} (\bibinfo {year} {2001})}\BibitemShut
  {NoStop}%
\bibitem [{\citenamefont {L{\'o}pez}\ \emph {et~al.}(2014)\citenamefont
  {L{\'o}pez}, \citenamefont {Lee}, \citenamefont {Serra},\ and\ \citenamefont
  {Lim}}]{Lopez2014PRB}%
  \BibitemOpen
  \bibfield  {author} {\bibinfo {author} {\bibfnamefont {R.}~\bibnamefont
  {L{\'o}pez}}, \bibinfo {author} {\bibfnamefont {M.}~\bibnamefont {Lee}},
  \bibinfo {author} {\bibfnamefont {L.}~\bibnamefont {Serra}}, \ and\ \bibinfo
  {author} {\bibfnamefont {J.~S.}\ \bibnamefont {Lim}},\ }\href {\doibase
  10.1103/PhysRevB.89.205418} {\bibfield  {journal} {\bibinfo  {journal}
  {Physical Review B}\ }\textbf {\bibinfo {volume} {89}},\ \bibinfo {pages}
  {205418} (\bibinfo {year} {2014})}\BibitemShut {NoStop}%
\bibitem [{\citenamefont {Ramos-Andrade}\ \emph {et~al.}(2016)\citenamefont
  {Ramos-Andrade}, \citenamefont {{\'A}valos-Ovando}, \citenamefont
  {Orellana},\ and\ \citenamefont {Ulloa}}]{Ramos-Andrade2016PRB}%
  \BibitemOpen
  \bibfield  {author} {\bibinfo {author} {\bibfnamefont {J.~P.}\ \bibnamefont
  {Ramos-Andrade}}, \bibinfo {author} {\bibfnamefont {O.}~\bibnamefont
  {{\'A}valos-Ovando}}, \bibinfo {author} {\bibfnamefont {P.~A.}\ \bibnamefont
  {Orellana}}, \ and\ \bibinfo {author} {\bibfnamefont {S.~E.}\ \bibnamefont
  {Ulloa}},\ }\href {\doibase 10.1103/PhysRevB.94.155436} {\bibfield  {journal}
  {\bibinfo  {journal} {Physical Review B}\ }\textbf {\bibinfo {volume} {94}},\
  \bibinfo {pages} {155436} (\bibinfo {year} {2016})}\BibitemShut {NoStop}%
\bibitem [{\citenamefont {Jiang}\ \emph {et~al.}(2012)\citenamefont {Jiang},
  \citenamefont {Entin-Wohlman},\ and\ \citenamefont {Imry}}]{Jiang2012PRB}%
  \BibitemOpen
  \bibfield  {author} {\bibinfo {author} {\bibfnamefont {J.-H.}\ \bibnamefont
  {Jiang}}, \bibinfo {author} {\bibfnamefont {O.}~\bibnamefont
  {Entin-Wohlman}}, \ and\ \bibinfo {author} {\bibfnamefont {Y.}~\bibnamefont
  {Imry}},\ }\href {\doibase 10.1103/PhysRevB.85.075412} {\bibfield  {journal}
  {\bibinfo  {journal} {Physical Review B}\ }\textbf {\bibinfo {volume} {85}},\
  \bibinfo {pages} {075412} (\bibinfo {year} {2012})}\BibitemShut {NoStop}%
\bibitem [{\citenamefont {Jiang}\ \emph {et~al.}(2013)\citenamefont {Jiang},
  \citenamefont {Entin-Wohlman},\ and\ \citenamefont {Imry}}]{Jiang2013NJP}%
  \BibitemOpen
  \bibfield  {author} {\bibinfo {author} {\bibfnamefont {J.-H.}\ \bibnamefont
  {Jiang}}, \bibinfo {author} {\bibfnamefont {O.}~\bibnamefont
  {Entin-Wohlman}}, \ and\ \bibinfo {author} {\bibfnamefont {Y.}~\bibnamefont
  {Imry}},\ }\href {\doibase 10.1088/1367-2630/15/7/075021} {\bibfield
  {journal} {\bibinfo  {journal} {New Journal of Physics}\ }\textbf {\bibinfo
  {volume} {15}},\ \bibinfo {pages} {075021} (\bibinfo {year}
  {2013})}\BibitemShut {NoStop}%
\bibitem [{\citenamefont {Mazza}\ \emph {et~al.}(2014)\citenamefont {Mazza},
  \citenamefont {Bosisio}, \citenamefont {Benenti}, \citenamefont
  {Giovannetti}, \citenamefont {Fazio},\ and\ \citenamefont
  {Taddei}}]{Mazza2014NJP}%
  \BibitemOpen
  \bibfield  {author} {\bibinfo {author} {\bibfnamefont {F.}~\bibnamefont
  {Mazza}}, \bibinfo {author} {\bibfnamefont {R.}~\bibnamefont {Bosisio}},
  \bibinfo {author} {\bibfnamefont {G.}~\bibnamefont {Benenti}}, \bibinfo
  {author} {\bibfnamefont {V.}~\bibnamefont {Giovannetti}}, \bibinfo {author}
  {\bibfnamefont {R.}~\bibnamefont {Fazio}}, \ and\ \bibinfo {author}
  {\bibfnamefont {F.}~\bibnamefont {Taddei}},\ }\href {\doibase
  10.1088/1367-2630/16/8/085001} {\bibfield  {journal} {\bibinfo  {journal}
  {New Journal of Physics}\ }\textbf {\bibinfo {volume} {16}},\ \bibinfo
  {pages} {085001} (\bibinfo {year} {2014})}\BibitemShut {NoStop}%
\bibitem [{\citenamefont {Jiang}\ \emph {et~al.}(2015)\citenamefont {Jiang},
  \citenamefont {Kulkarni}, \citenamefont {Segal},\ and\ \citenamefont
  {Imry}}]{Jiang2015PRB}%
  \BibitemOpen
  \bibfield  {author} {\bibinfo {author} {\bibfnamefont {J.-H.}\ \bibnamefont
  {Jiang}}, \bibinfo {author} {\bibfnamefont {M.}~\bibnamefont {Kulkarni}},
  \bibinfo {author} {\bibfnamefont {D.}~\bibnamefont {Segal}}, \ and\ \bibinfo
  {author} {\bibfnamefont {Y.}~\bibnamefont {Imry}},\ }\href {\doibase
  10.1103/PhysRevB.92.045309} {\bibfield  {journal} {\bibinfo  {journal}
  {Physical Review B}\ }\textbf {\bibinfo {volume} {92}},\ \bibinfo {pages}
  {045309} (\bibinfo {year} {2015})}\BibitemShut {NoStop}%
\bibitem [{\citenamefont {Whitney}\ \emph {et~al.}(2018)\citenamefont
  {Whitney}, \citenamefont {S{\'a}nchez},\ and\ \citenamefont
  {Splettstoesser}}]{Whitney2018book}%
  \BibitemOpen
  \bibfield  {author} {\bibinfo {author} {\bibfnamefont {R.~S.}\ \bibnamefont
  {Whitney}}, \bibinfo {author} {\bibfnamefont {R.}~\bibnamefont
  {S{\'a}nchez}}, \ and\ \bibinfo {author} {\bibfnamefont {J.}~\bibnamefont
  {Splettstoesser}},\ }\enquote {\bibinfo {title} {Quantum thermodynamics of
  nanoscale thermoelectrics and electronic devices},}\ in\ \href {\doibase
  10.1007/978-3-319-99046-0_7} {\emph {\bibinfo {booktitle} {Thermodynamics in
  the Quantum Regime: Fundamental Aspects and New Directions}}},\ \bibinfo
  {editor} {edited by\ \bibinfo {editor} {\bibfnamefont {F.}~\bibnamefont
  {Binder}}, \bibinfo {editor} {\bibfnamefont {L.~A.}\ \bibnamefont {Correa}},
  \bibinfo {editor} {\bibfnamefont {C.}~\bibnamefont {Gogolin}}, \bibinfo
  {editor} {\bibfnamefont {J.}~\bibnamefont {Anders}}, \ and\ \bibinfo {editor}
  {\bibfnamefont {G.}~\bibnamefont {Adesso}}}\ (\bibinfo  {publisher} {Springer
  International Publishing},\ \bibinfo {address} {Cham},\ \bibinfo {year}
  {2018})\ pp.\ \bibinfo {pages} {175--206}\BibitemShut {NoStop}%
\bibitem [{\citenamefont {Blundell}\ and\ \citenamefont
  {Blundell}(2010)}]{blundell2010}%
  \BibitemOpen
  \bibfield  {author} {\bibinfo {author} {\bibfnamefont {S.~J.}\ \bibnamefont
  {Blundell}}\ and\ \bibinfo {author} {\bibfnamefont {K.~M.}\ \bibnamefont
  {Blundell}},\ }\href {\doibase 10.1093/acprof:oso/9780199562091.001.0001}
  {\emph {\bibinfo {title} {Concepts in thermal physics}}}\ (\bibinfo
  {publisher} {Oxford University Press},\ \bibinfo {year} {2010})\BibitemShut
  {NoStop}%
\bibitem [{Note3()}]{Note3}%
  \BibitemOpen
  \bibinfo {note} {The relation between $\Gamma _{\alpha }$ and $\lambda
  _{\alpha j}$ can be found in our previous paper \protect \citep
  {Zhang2021pre}}\BibitemShut {NoStop}%
\end{thebibliography}%

\end{document}